\documentclass[11pt,graphicx]{article}
\usepackage{amssymb,amsmath,amsfonts}
\usepackage{graphicx}
\usepackage{graphics}
\usepackage{eepic,epsfig}

\@addtoreset{equation}{section}

\makeatother

\begin{document}

\title{Scalar self-energy for a charged particle in global monopole
spacetime with a spherical boundary}
\author{E. R. Bezerra de Mello$^{1}$\thanks{%
E-mail: emello@fisica.ufpb.br}\, and A. A. Saharian$^{1,2}$\thanks{%
E-mail: saharian@ysu.am} \\
\\
\textit{$^1$Departamento de F\'{\i}sica-CCEN, Universidade Federal da Para
\'{\i}ba}\\
\textit{58.059-970, Caixa Postal 5.008, Jo\~{a}o Pessoa, PB, Brazil}\vspace{%
0.3cm}\\
\textit{$^2$Department of Physics, Yerevan State University,}\\
\textit{1 Alex Manoogian Street, 0025 Yerevan, Armenia}}
\maketitle

\begin{abstract}
We analyze combined effects of the geometry produced by global monopole and
a concentric spherical boundary on the self-energy of a point-like scalar
charged test particle at rest. We assume that the boundary is outside the
monopole's core with a general spherically symmetric inner structure. An
important quantity to this analysis is the three-dimensional Green function
associated with this system. For both Dirichlet and Neumann boundary
conditions obeyed by the scalar field on the sphere, the Green function
presents a structure that contains contributions due to the background
geometry of the spacetime and the boundary. Consequently the corresponding
induced scalar self-energy present also similar structure. For points near
the sphere the boundary-induced part dominates and the self-force is
repulsive/attractive with respect to the boundary for Dirichlet/Neumann
boundary condition. In the region outside the sphere at large distances from
it, the boundary-free part in the self-energy dominates and the
corresponding self-force can be either attractive or repulsive with
dependence of the curvature coupling parameter for scalar field. In
particular, for the minimal coupling we show the presence of a stable
equilibrium point for Dirichlet boundary condition. In the region inside the
sphere the nature of the self-force depends on the specific model for the
monopole's core. As illustrations of the general procedure adopted we shall
consider two distinct models, namely flower-pot and the ballpoint-pen ones.
\end{abstract}

\bigskip


\bigskip

\section{Introduction}

\label{Int}

It is well known that different types of topological objects may have been
formed in the early Universe by vacuum phase transitions after Planck time
\cite{Kibble,V-S}. These include domain walls, cosmic strings and monopoles.
Global monopoles are heavy and spherically symmetric topological defects.
The simplest theoretical model where global monopoles can be obtained is
given by a Lagrange density composed by self-coupling iso-triplet scalar
field which presents spontaneous breakdown of global $O(3)$ gauge symmetry
\cite{BV}. For points far away from the monopole's center, the geometry of
the spacetime can be given by the line element below,
\begin{equation}
ds^{2}=-dt^{2}+dr^{2}+\alpha ^{2}r^{2}(d\theta ^{2}+\sin ^{2}\theta d\varphi
^{2})\ ,  \label{gm}
\end{equation}%
where the parameter $\alpha $, smaller than unity, depends on the energy
scale where the monopole is formed. The spacetime described by (\ref{gm})
has a non-vanishing scalar curvature, $R=2(\alpha ^{-2}-1)/r^{2}$, and
presents a solid angle deficit $\delta \Omega =4\pi ^{2}(1-\alpha ^{2})$.

Although the geometric properties of the spacetime outside the monopole are
very well known, no explicit expressions for the components of the metric
tensor in the region inside have been found. Consequently many interesting
investigations associated with global monopole consider this object as a
point-like defect. In this way calculations of vacuum polarization effects
associated with bosonic \cite{Lousto} and fermionic quantum fields \cite%
{Mello2}, in four-dimensional spacetime, present divergence at the
monopole's center. Moreover, considering higher-dimensional spacetime,
vacuum polarization effects associated with bosonic \cite{Mello2a} and
fermionic \cite{Mello2b} quantum fields, also present similar divergences.

Many years ago, Linet \cite{Linet} and Smith \cite{Smith}, independently,
have shown that an electrically charged particle placed at rest in the
spacetime of an idealized cosmic string becomes subjected to a repulsive
self-interaction. This self-interaction is a consequence of the distortion
of the particle's fields caused by the planar angle deficit associated with
the conical geometry. Also it has been shown in \cite{Furtado} that a linear
electric or magnetic sources in the spacetime of a cosmic string parallel to
the latter, become subject to induced self-interactions. In the spacetime of
a global monopole the phenomenon of induced electrostatic self-interaction
has been analyzed in \cite{Mello3}, considering the defect as a point-like
object. In all these analysis, the corresponding self-forces present
divergences on the respective defects' core. A possible way to circumvent
the divergence problem is to consider these defects as having a
non-vanishing radius, and attributing for the region inside a structure. For
the cosmic string, two different models have been adopted to describe the
geometry inside it. The ballpoint pen model proposed independently by Gott
and Hiscock \cite{Gott}, replaces the conical singularity at the string axis
by a constant curvature space in the interior region, and the flower-pot
model \cite{BA}, presents the curvature concentrated on a ring with the
spacetime inside the string been flat. In \cite{NV} the problem of
electrostatic self-energy in cosmic string spacetime has been revisited by
considering the Gott and Hiscock model for the region inside. Concerning to
the global monopole spacetime, the electrostatic self-energy problem has
been analyzed considering for the region inside, the flower-pot and
ballpoint pen models in \cite{Mello4} and \cite{Barbosa,Barbosa1},
respectively. For both defects it was observed that the corresponding
self-forces are finite at their center.\footnote{%
The analysis of vacuum polarization effect associated with scalar and
fermionic fields, considering the flower-pot model for the region inside the
monopole, have been developed in \cite{Mello5} and \cite{Mello6},
respectively.}

Differently from the electrostatic analysis, the induced self-energy on
scalar charged point-like particles on a curved spacetime presents
peculiarities due to the non-minimal curvature coupling with the geometry
\cite{Burko,Wiseman}. In the case of of Schwarzschild and Kerr spacetimes,
the scalar self-force on a charged particle have been considered in \cite%
{Zelnikov} and \cite{Barack}, respectively. Also, the self-energy on scalar
particle in the wormhole spacetime has been developed recently in \cite{NV1}%
. In the latter, the authors have observed that the self-force vanishes for
a conformally coupled massless field. Recently the induced self-energy on a
static scalar charged particle in the global monopole has been analyzed in
\cite{Barbosa2} considering a inner structure to it.

In many problems we need to consider a field-theoretical model on the
background of manifolds with boundaries on which the dynamical variables
satisfy some prescribed boundary conditions. Combined effects of the
geometry produced by the global monopole and concentric spherical boundaries
on the polarization of scalar vacuum have been investigated in \cite%
{Saha03,Saha04}. Here we decided to continue the analysis of
induced self-energies on a static scalar point-like charged
particle in the global monopole spacetime, considering at this
time the presence of a spherical boundary concentric with the
monopole's center, and imposing on the field Dirichlet and Neumann
boundary conditions. Following the same line of investigation
developed in \cite{Barbosa2}, we consider an inner structure for
the monopole; moreover we adopt the core's radius of the monopole,
$a$, being smaller that the boundary's radius, $R$. In this
analysis two distinct situations are taken into account: the
charged particle placed at rest outside the boundary followed by
the particle inside the boundary. In the latter two different
configurations are separately treated, the particle outside the
core and inside the core.

Because the analysis is first developed considering a general spherically
symmetric structure for the region inside the monopole, two different
applications of this formalism are considered, adopting the flower-pot and
ballpoint pen models for the region inside. The organization of this paper
is the following. In section \ref{system} we present our approach to
consider the geometry of the spacetime under consideration, the relevant
field equations associated with the scalar charged particle and the boundary
conditions obeyed by this field. Also we present the Green functions
corresponding to all physical situations. In section \ref{bc}, explicit
formulas for the self-energies ere presented, considering the particle
outside the boundary and inside the boundary. As illustrations of the
general analysis, in section \ref{Applic} the flower-pot and ballpoint pen
models for the region inside the monopole's core are considered separately.
Because the scalar self-energy in the global monopole spacetime has already
been investigated, here we are mainly interested in the contribution induced
by the boundary. Moreover, we analyze the behavior of the self-energy in
various asymptotic regions. In section \ref{Conc}, we present our
conclusions and more relevant remarks. We leave for the appendix \ref{appA}
specific details related with the obtainment of a simple expression for the
energy, and the conditions that the field must obey on the spherical
boundary. In the appendix \ref{appB} we provide the procedure adopted to the
development of numerical analysis of the self-energy.

\section{The system}

\label{system} In this section we present our approach to investigate the
induced scalar self-energy, providing a general spherically symmetric model
to describe the metric tensor in the region inside the monopole. Also the
field equations and boundary conditions obeyed by the field are presented,
in order to provide specific expressions for the Green functions considering
different situations related with the position of the particle.

\subsection{The model}

Many investigations of the physical effects associated with a global
monopole are developed considering it as an idealized point-like defect. In
these analysis the geometry of the whole spacetime is described by line
element (\ref{gm}). However, a realistic model for a global monopole has to
take into account a non-vanishing core radius. Specifically in the system
proposed by Barriola and Vilenkin \cite{BV}, the line element given by (\ref%
{gm}) is attained for the radial coordinate much larger than its
characteristic core radius, which depends on the inverse of the energy scale
where the global $O(3)$ symmetry is spontaneously broken to $U(1)$. However,
explicit expressions for the components of the metric tensor in whole space
have not yet been found. Here, in this paper we shall not go into the
details about this unsolved problem. Instead, we shall consider a simplified
model described by two sets of the metric tensor for two different regions,
continuous at a spherical shell of radius $a$. In the exterior region
corresponding to $r>a$, the line element is given by (\ref{gm}), while in
the interior region, $r<a$, the geometry is described by the static
spherically symmetric line element
\begin{equation}
ds^{2}=-dt^{2}+v^{2}(r)dr^{2}+w^{2}(r)(d\theta ^{2}+\sin ^{2}\theta d\varphi
^{2})\ .  \label{gm1}
\end{equation}%
As the metric tensor must be continuous at the core's boundary, the
functions $v(r)$ and $w(r)$ must satisfy the conditions
\begin{equation}
v(a)=1\ \mathrm{and}\ w(a)=\alpha a\ .  \label{bound}
\end{equation}%
The flower-pot and ballpoint pen models for a global monopole, can be
implemented from the general expression given in (\ref{gm1}) by taking
\begin{equation}
v(r)=1\ ,\ w(r)=r+(\alpha -1)a  \label{FP}
\end{equation}%
for the flower-pot model and
\begin{equation}
v(r)=\alpha \left[ 1-(r/a)^{2}(1-\alpha ^{2})\right] ^{-1/2}\ ,\ w(r)=\alpha
r\ ,  \label{BP}
\end{equation}%
for the ballpoint pen one.

\subsection{Scalar self-energy}
In this section we analyze a massive scalar field, $\phi $, generated by a scalar
charge density, $\rho $, in background of the geometry described
by the line element (\ref{gm1}). In the problem under consideration the 
particle with the scalar charge is considered as an extra
degree of freedom with the fixed location, so its dynamics 
is not taking into account. In addition, we assume
the presence of a spherical boundary, concentric with the global
monopole, on which the field obeys Dirichlet or Neumann boundary
conditions. We shall discuss the both cases with the charge
location outside and inside the spherical boundary. Due to the
distortion of the particle's field caused by the geometry of a
global monopole and boundary conditions obeyed by the field, a
self-action force appears acting on the particle. In this paper we
shall consider the influence of a spherical boundary and the monopole's
non-trivial core structure on this force by evaluating the
corresponding self-energy.

The action associated with a massive scalar field, $\phi$, coupled with a
scalar charge density, $\rho$, in a curved background spacetime can be given
by
\begin{equation}  \label{Action}
S=-\frac12\int \ d^4x \ \sqrt{-g} \
\left(g^{\mu\nu}\nabla_\mu\phi\nabla_\nu\phi +\xi {\mathcal{R}}%
\phi^2+m^2\phi^2\right) +\int \ d^4x \ \sqrt{-g} \ \rho \ \phi \ ,
\end{equation}
where the first part corresponds to the Klein-Gordon action with an
arbitrary curvature coupling, $\xi$, and the second part takes into account
the presence of an interaction term. Note that this type of interaction is
used in the Unruh-DeWitt model for a particle detector (see \cite{B-D}). In
the above equation ${\mathcal{R}}$ represents the scalar curvature and $g$
the determinant of $g_{\mu\nu}$.

The field equation obtained by varying the action with respect to the field
reads,
\begin{equation}
\left( \Box -\xi {\mathcal{R}}-m^{2}\right) \phi =-\rho \ .  \label{EM}
\end{equation}%
Because the physical system which we are interested corresponds to a static
charged particle, the field is time independent, consequently the equation (%
\ref{EM}) reduces to an effective three-dimensional field equation below:
\begin{equation}
\left( \nabla ^{2}-\xi {\mathcal{R}}-m^{2}\right) \phi =-\rho \ .
\label{EM1}
\end{equation}

Taking the variation of (\ref{Action}) with respect to the metric tensor we
obtain the energy-momentum tensor associated with this system. For the
further discussion it is convenient to write the corresponding expression in
the form%
\begin{eqnarray}
T_{\mu \nu } &=&\nabla _{\mu }\phi \nabla _{\nu }\phi +\frac{1}{2}g_{\mu \nu
}\phi \left( \Box -m^{2}-\xi {\mathcal{R}}\right) \phi  \notag \\
&&+\left[ (\xi -1/4)g_{\mu \nu }\Box +\xi R_{\mu \nu }-\xi \nabla _{\mu
}\nabla _{\nu }\right] \phi ^{2}+\rho \ \phi \ g_{\mu \nu }\ ,  \label{Tmunu}
\end{eqnarray}
with $R_{\mu \nu }$ being the Ricci tensor.

The energy associated with the scalar particle is given by,
\begin{equation}
E=-\int \ d^{3}x\ \sqrt{-g}\ T_{0}^{0}\ .  \label{Energy1}
\end{equation}%
(Note that the energy considered in \cite{Barbosa2} is the 0-th covariant
component of the four-momentum and differs by sign from (\ref{Energy1})). In
appendix \ref{appA} we show that, considering a spherical boundary with
radius $R$ on background of the manifold described by the line element (\ref%
{gm1}), this energy may be presented in the form
\begin{equation}
E=-\frac{1}{2}\int \ d^{3}x\ \sqrt{-g}\ \rho \ \phi \ ,  \label{SE}
\end{equation}%
for fields obeying, separately, Dirichlet
\begin{equation}
\phi (r=R,\theta ,\varphi )=0\ ,  \label{Diri1}
\end{equation}%
or Neumann
\begin{equation}
\partial _{r}\phi (r,\theta ,\varphi )|_{r=R}=0\ ,  \label{Neu1}
\end{equation}%
conditions on the boundary. In (\ref{SE}) the integral is evaluated on the
inner, $r\leqslant R$, or outer, $r\geqslant R$, regions, for the case of
the particle inside or outside the boundary.

Considering now a point-like scalar charge at rest at the point $\mathbf{x}%
_{p}$, the charge density takes the form,
\begin{equation}
\rho (\mathbf{x})=\frac{q}{\sqrt{-g}}\delta ^{3}(\mathbf{x}-\mathbf{x}_{p})\
.  \label{Charge}
\end{equation}%
For this case, the scalar field and the effective three-dimensional Green
function defined below,
\begin{equation}
\left( \nabla ^{2}-\xi \mathcal{R}-m^{2}\right) G(\mathbf{x},\mathbf{x}%
^{\prime })=-\frac{\delta ^{3}(\mathbf{x}-\mathbf{x}^{\prime })}{\sqrt{-g}}\
,  \label{Green}
\end{equation}%
are related by
\begin{equation}
\phi (\mathbf{x})=q\ G(\mathbf{x},\mathbf{x}_{p})\ .
\end{equation}%
So, \eqref{SE} becomes,
\begin{equation}
E=-\frac{q^{2}}{2}G(\mathbf{x}_{p},\mathbf{x}_{p})\ .  \label{E2}
\end{equation}

The effective three-dimensional Green function that we need for the
calculation of the energy diverges at the coincidence limit, providing also
a divergent result for the energy. So, we have to apply some renormalization
approach to obtain a finite and well defined result. Here we shall use a
general renormalization approach in curved spacetime \cite{B-D}, which
consists to subtract from the Green function its DeWitt-Schwinger asymptotic
expansion. In general there are two types of divergences in the expansion,
namely, pole and logarithmic ones. In three-dimensional case which we are
interested in, there is only pole divergence. Following \cite{Chris}, the
DeWitt-Schwinger expansion for the three-dimensional Green function
associated with a massive scalar field reads,
\begin{equation}
G_{\mathrm{Sing}}(\mathbf{x},\mathbf{x}^{\prime })=\frac{1}{4\pi }\left(
\frac{1}{\sqrt{2\sigma }}-m\right) +O(\sigma )\ ,  \label{Had3D}
\end{equation}%
where $\sigma $ is the one-half of the geodesic distance between two points.
Adopting the renormalization approach for curved space, the renormalized
scalar self-energy is given by
\begin{equation}
E_{\mathrm{Ren}}=-\frac{q^{2}}{2}\lim_{\mathbf{x}\rightarrow \mathbf{x}%
_{p}}[G(\mathbf{x},\mathbf{x}_{p})-G_{\mathrm{Sing}}(\mathbf{x},\mathbf{x}%
_{p})]\,.  \label{ERen}
\end{equation}

Having the self-energy, we can evaluate the corresponding self-force by
using the formula%
\begin{equation}
\mathbf{F}_{\mathrm{Ren}}=-\mathbf{\nabla }_{\mathbf{x}_{p}}E_{\mathrm{Ren}%
}\,.  \label{F1}
\end{equation}%
An alternative form is obtained by employing the expression of the force $%
\mathbf{F}=q\mathbf{\nabla }_{\mathbf{x}}\phi (\mathbf{x})|_{\mathbf{x}=%
\mathbf{x}_{p}}$. For the renormalized self-force this gives%
\begin{equation}
\mathbf{F}_{\mathrm{Ren}}=q^{2}\lim_{\mathbf{x}\rightarrow \mathbf{x}_{p}}%
\mathbf{\nabla }_{\mathbf{x}}[G(\mathbf{x},\mathbf{x}_{p})-G_{\mathrm{Sing}}(%
\mathbf{x},\mathbf{x}_{p})].  \label{F2}
\end{equation}%
By the Synge's theorem, two expressions (\ref{F1}) and (\ref{F2}) are
equivalent.

\subsection{Green function}

Taking into account the spherical symmetry of the problem, the scalar Green
function can be expressed by the following ansatz,
\begin{equation}
G(\mathbf{x},\mathbf{x}^{\prime })=\sum_{l=0}^{\infty
}\sum_{m=-l}^{l}g_{l}(r,r^{\prime })Y_{l}^{m}(\theta ,\varphi )Y_{l}^{m\ast
}(\theta ^{\prime },\varphi ^{\prime })\ ,  \label{Green-a}
\end{equation}%
with $Y_{l}^{m}(\theta ,\varphi )$ being the ordinary spherical harmonics.
Because here we are analyzing the problem related with scalar charged field
in a global monopole spacetime in the presence of a spherical boundary of
radius $R$, we have to solve the differential equation for the Green
function, $G(\mathbf{x},\mathbf{x}^{\prime })$, by imposing on this function
each of the boundary conditions below:
\begin{equation}
G(r,\theta ,\varphi ;\ r^{\prime },\theta ^{\prime },\varphi ^{\prime })=0\
\mathrm{for}\ r=R\ ,
\end{equation}%
for Dirichlet boundary condition (DBC), or
\begin{equation}
\partial _{r}G(r,\theta ,\varphi ;\ r^{\prime },\theta ^{\prime },\varphi
^{\prime })=0\ \mathrm{for}\ r=R\ ,
\end{equation}%
for Neumann boundary condition (NBC). In the above expressions, $r$
corresponds to the radial distance of the point closer to the boundary.

In the following calculations we shall adopt a compact notation below,
\begin{equation}
\left( A+B\partial _{r}\right) G(\mathbf{x},\mathbf{x}^{\prime })=0\ \mathrm{%
for}\ r=R\ ,  \label{DNC}
\end{equation}%
and at the end, we specify the values of the parameters associated with each
conditions, i.e., $B=0$ for DBC and $A=0$ for NBC. Moreover, as it have been
mentioned before, we shall assume that the boundary is outside the
monopole's core, $R>a$.

In what follows we shall present the steps needed for the calculation of the
Green function considering the general expression for the metric tensor (\ref%
{gm1}). Substituting (\ref{Green-a}) into (\ref{Green}) and using the
closure relation for the spherical harmonics, we obtain the radial
differential equation below obeyed by the unknown function $%
g_{l}(r,r^{\prime })$:
\begin{equation}
\left[ \frac{d}{dr}\left( \frac{w^{2}}{v}\frac{d}{dr}\right) -l(l+1)v-\xi {%
\mathcal{R}}vw^{2}-m^{2}vw^{2}\right] g_{l}(r,r^{\prime })=-\delta
(r-r^{\prime })\ ,  \label{gr}
\end{equation}%
with the Ricci scalar being given by
\begin{equation}
{\mathcal{R}}=\frac{2}{w^{2}}+\frac{4v^{\prime }w^{\prime }}{wv^{3}}-\frac{%
4w^{\prime \prime }}{wv^{2}}-\frac{2(w^{\prime })^{2}}{w^{2}v^{2}}\ .
\end{equation}

At this point we would like to analyze separately two distinct situations.
The first one corresponds to the particle outside the boundary, $%
r_{p}=r^{\prime }>R$, and the second to the particle inside the boundary, $%
r_{p}=r^{\prime }<R$.

\subsubsection{Particle outside the boundary}

Because we are considering $R>a$, in this first analysis the metric tensor
is given by (\ref{gm}) and the Ricci scalar reads,
\begin{equation}
{\mathcal{R}}=2\frac{1-\alpha ^{2}}{\alpha ^{2}r^{2}}\ ,
\end{equation}%
consequently (\ref{gr}) becomes
\begin{equation}
\left[ \frac{d}{dr}\left( r^{2}\frac{d}{dr}\right) -\frac{l(l+1)+2\xi
(1-\alpha ^{2})}{\alpha ^{2}}-m^{2}r^{2}\right] g_{l}(r,r^{\prime })=-\frac{%
\delta (r-r^{\prime })}{\alpha ^{2}}\ .  \label{gr0}
\end{equation}%
The function $g_{l}(r,r^{\prime })$ is continuous at $r=r^{\prime }$,
however by integrating (\ref{gr0}) about this point, the first radial
derivative of this function obeys the following junction condition:
\begin{equation}
\partial _{r}g_{l}(r,r^{\prime })|_{r=r^{\prime }{}^{+}}-\partial
_{r}g_{l}(r,r^{\prime })|_{r=r^{\prime }{}^{-}}=-\frac{1}{\alpha
^{2}r^{\prime }{}^{2}}\ .  \label{gr1}
\end{equation}

In the region under consideration the two linearly independent solutions of
the homogeneous equation corresponding to (\ref{gr0}) are:
\begin{equation}
\frac{1}{\sqrt{r}}I_{\nu _{l}}(mr)\ \ \mathrm{and}\ \frac{1}{\sqrt{r}}K_{\nu
_{l}}(mr)\ ,  \label{ext-sol}
\end{equation}%
where $I_{\nu }$ and $K_{\nu }$ are the modified Bessel functions of the
order
\begin{equation}
\nu _{l}=\frac{1}{2\alpha }\sqrt{(2l+1)^{2}+(1-\alpha ^{2})(8\xi -1)}\ .
\label{ind}
\end{equation}%
Assuming that for $r>r^{\prime }$, $g_{l}(r,r^{\prime })$, defined below by $%
g_{l}^{>}(r,r^{\prime })$, goes to zero at infinity, we take for this region
\begin{equation}
g_{l}^{>}(r,r^{\prime })=\frac{a_{l}}{\sqrt{r}}K_{\nu _{l}}(mr)\ .
\end{equation}%
For the region $R<r<r^{\prime }$, the radial function, defined as $%
g_{l}^{<}(r,r^{\prime })$, takes the form:
\begin{equation}
g_{l}^{<}(r,r^{\prime })=\frac{b_{l}}{\sqrt{r}}K_{\nu _{l}}(mr)+\frac{c_{l}}{%
\sqrt{r}}I_{\nu _{l}}(mr)\ .
\end{equation}%
In the above expression the term with the Macdonald function is a
consequence of the presence of the boundary. Following the compact notation
specified in (\ref{DNC}), DBC or NBC imposed on $g_{l}^{<}(r,r^{\prime })$
can be written as:
\begin{equation}
\left( A+B\partial _{r}\right) g_{l}^{<}(r,r^{\prime })=0 \ \mathrm{at}\ r=R\ .
\label{DNC1}
\end{equation}%
Finally using the junction condition (\ref{gr1}) together with (\ref{DNC1}),
the expression for the radial function $g_{l}$ reads:
\begin{equation}
g_{l}(r,r^{\prime })=\frac{I_{\nu _{l}}(mr_{<})K_{\nu _{l}}(mr_{>})}{\alpha
^{2}\sqrt{rr^{\prime }}}-\frac{K_{\nu _{l}}(mr)K_{\nu _{l}}(mr^{\prime })}{%
\alpha ^{2}\sqrt{rr^{\prime }}}\frac{\bar{I}_{\nu _{l}}(mR)}{\bar{K}_{\nu
_{l}}(mR)}\ ,  \label{goutR}
\end{equation}%
where $r_{<}=\min (r,r^{\prime })$ and $r_{>}=\max (r,r^{\prime })$, and we
have used the notations,
\begin{equation}
\bar{F}(z)=ARF(z)+B\tilde{F}(z)\ ,\quad \mathrm{with}\quad \tilde{F}%
(z)=zF^{\prime }(z)-\frac{1}{2}F(z)\ .  \label{F-bar}
\end{equation}%
In (\ref{F-bar}), $F^{\prime }(z)$ denotes the derivative of $F(z)$ with
respect to its argument. The second term in the right-hand side of (\ref%
{goutR}) goes to zero for $mR\rightarrow 0$; also we can verify that $g_{l}$
or $\partial _{r_{<}}g_{l}$ go to zero for $r_{<}=R$ for DBC and NBC,
respectively

\subsubsection{Particle inside the boundary}

Now in this second analysis we shall consider the particle inside the
boundary. In this case two sub cases are present: The particle outside the
monopole's core and the particle inside the core. Here it is more
appropriate to use for the metric tensor the expression (\ref{gm1}). Because
the functions $v(r)$ and $w(r)$ are continuous at $r=a$, from \eqref{gr} it
follows that $g_{l}(r,r^{\prime })$ should be continuous at this point.
However, due to the second radial derivative of the function $w(r)$ in the
expression for the Ricci scalar, a Dirac-delta function contribution appears
if the first derivative of this function is not continuous at the monopole's
core.\footnote{%
For the flower-pot model this fact occurs and has been considered in \cite%
{Mello4}.} Naming by $\check{R}=\bar{R}\delta (r-a)$ the Dirac-delta
contribution of the Ricci scalar, the junction condition on the monopole's
core is:
\begin{equation}
\partial _{r}g_{l}(r,r^{\prime })|_{r=a^{+}}-\partial _{r}g_{l}(r,r^{\prime
})|_{r=a^{-}}=\xi \bar{R}\ g_{l}(a,r^{\prime })\ .  \label{Cond1}
\end{equation}%
Moreover, the function $g_{l}(r,r^{\prime })$ is continuous at $r=r^{\prime
} $, however by integrating (\ref{gr}) about this point, the first radial
derivative of this function obeys the junction condition below,
\begin{equation}
\partial _{r}g_{l}(r,r^{\prime })|_{r=r^{\prime +}}-\partial
_{r}g_{l}(r,r^{\prime })|_{r=r^{\prime -}}=-\frac{v(r^{\prime })}{%
w^{2}(r^{\prime })}\ .  \label{Cond2}
\end{equation}%
Finally we must impose on this function DBC or NBC. These conditions can be
implemented as before:
\begin{equation}
\left( A+B\partial _{r}\right) g_{l}(r,r^{\prime })=0\ \mathrm{at}\ r=R\ .
\label{DNB2}
\end{equation}

Now after this general discussion, let us analyze the solutions of the
homogeneous differential equation associated with (\ref{gr}) for regions
inside the monopole's core. Denoting by $R_{1l}(r)$ and $R_{2l}(r)$ two
linearly independent solutions of this equation, we shall assume that $%
R_{1l}(r)$ is regular at the core center $r=r_{c}$ and that the solutions
are normalized by the Wronskian relation
\begin{equation}
R_{1l}(r)R_{2l}^{\prime }(r)-R_{1l}^{\prime }(r)R_{2l}(r)=-\frac{v(r)}{%
w^{2}(r)}.  \label{Wronin}
\end{equation}%
Moreover, for the region outside the core the solutions are given by %
\eqref{ext-sol} and \eqref{ind}.

Now we can write the function $g_{l}(r,r^{\prime })$ as a linear combination
of the above solutions with arbitrary coefficients for the regions $%
(r_{c},\min (r^{\prime },a))$, $(\min (r^{\prime },a),\ \max (r^{\prime },a))
$, and $(\max (r^{\prime },a),R)$. The requirement of the regularity at the
core center and the boundary condition, reduce the number of these
coefficients to five. These constants are determined by the continuity
condition at the monopole's core and at the point $r=r^{\prime }$, by the
junctions conditions given in (\ref{Cond1}) and (\ref{Cond2}), respectively,
together with the boundary condition (\ref{DNB2}). In this way we find the
following expressions:
\begin{eqnarray}
g_{l}(r,r^{\prime }) &=&\frac{R_{1l}(r)K_{\nu _{l}}(mr^{\prime })}{\alpha
^{2}\sqrt{ar^{\prime }}}\frac{M_{\nu _{l}}(ma)}{R_{1l}(a)}\left[ 1-\frac{%
M_{\nu _{l}}(mr^{\prime })\bar{K}_{\nu _{l}}(mR)}{K_{\nu _{l}}(mr^{\prime })%
\bar{M}_{\nu _{l}}(mR)}\right] \ \ \mathrm{for}\ r\leqslant a\ ,
\label{gout-} \\
g_{l}(r,r^{\prime }) &=&\frac{I_{\nu _{l}}(mr_{<})K_{\nu _{l}}(mr_{>})}{%
\alpha ^{2}\sqrt{rr^{\prime }}}+D_{l}^{(+)}(a)\frac{K_{\nu _{l}}(mr)K_{\nu
_{l}}(mr^{\prime })}{\alpha ^{2}\sqrt{rr^{\prime }}}  \notag \\
&-&\frac{M_{\nu _{l}}(mr)M_{\nu _{l}}(mr^{\prime })}{\alpha ^{2}\sqrt{%
rr^{\prime }}}\frac{\bar{K}_{\nu _{l}}(mR)}{\bar{M}_{\nu _{l}}(mR)}\ \
\mathrm{for}\ r\geqslant a\ ,  \label{gout+}
\end{eqnarray}%
in the case of the charged particle outside the core, $r^{\prime }\geqslant a
$ , and
\begin{eqnarray}
g_{l}(r,r^{\prime })
&=&R_{1l}(r_{<})[R_{2l}(r_{>})+D_{l}^{(-)}(a)R_{1l}(r_{>})]\ \ \mathrm{for}\
r\leqslant a\ ,  \label{gin-} \\
g_{l}(r,r^{\prime }) &=&\frac{R_{1l}(r^{\prime })}{\alpha ^{2}\sqrt{ar}}%
\frac{1}{h_{l}^{(1)}(a)}\left[ K_{\nu _{l}}(mr)\bar{I}_{\nu _{l}}(mR)-I_{\nu
_{l}}(mr)\bar{K}_{\nu _{l}}(mR)\right] \ \ \mathrm{for}\ \ r\geqslant a\ ,
\label{gin+}
\end{eqnarray}%
for the charged particle inside the core, $r^{\prime }\leqslant a$. In these
formulas, $r_{<}=\min (r,r^{\prime })$ and $r_{>}=\max (r,r^{\prime })$, and
we have used the notations:
\begin{eqnarray}
M_{\nu }(z) &=&I_{\nu }(z)+D_{l}^{(+)}(a)K_{\nu }(z)\ ,  \label{M} \\
D_{l}^{(+)}(a) &=&-\frac{a{\mathcal{R}}_{l}^{(1)}(a)I_{\nu
_{l}}(ma)-R_{1l}(a)\tilde{I}_{\nu _{l}}(ma)}{a{\mathcal{R}}%
_{l}^{(1)}(a)K_{\nu _{l}}(ma)-R_{1l}(a)\tilde{K}_{\nu _{l}}(ma)}  \label{D+}
\end{eqnarray}%
and
\begin{equation}
D_{l}^{(-)}(a)=-\frac{h_{l}^{(2)}(a)}{h_{l}^{(1)}(a)}\ ,  \label{D-}
\end{equation}%
with
\begin{eqnarray}
h_{l}^{(j)}(a) &=&\left[ aK_{\nu _{l}}(ma){\mathcal{R}}_{l}^{(j)}(a)-\tilde{K%
}_{\nu _{l}}(ma)R_{jl}(a)\right] \bar{I}_{\nu _{l}}(mR)  \notag \\
&-&\left[ aI_{\nu _{l}}(ma){\mathcal{R}}_{l}^{(j)}(a)-\tilde{I}_{\nu
_{l}}(ma)R_{jl}(a)\right] \bar{K}_{\nu _{l}}(mR)\ .
\end{eqnarray}%
Here for any function $F(z)$ the notation for $\tilde{F}(z)$ was defined in %
\eqref{F-bar}, and for a solution $R_{jl}(r)$, with $j=1,\ 2$,
\begin{equation}
{\mathcal{R}}_{l}^{(j)}(r)=R_{jl}^{\prime }(r)+\xi \bar{R}R_{jl}(r)\ .
\label{Rsing}
\end{equation}%
The above expressions for $g_{l}$ are reduced to the previously obtained
ones in \cite{Barbosa2} for $mR\rightarrow \infty $. The functions (\ref%
{gout-}) and (\ref{gout+}) obey DBC or NBC for $r^{\prime }=R$ and for $%
r_{>}=R$, respectively. The same conditions are obeyed by (\ref{gin+}) for $%
r=R$.

\section{Self-energy}

\label{bc}

In this section we shall calculate the renormalized scalar self-energy
considering two situations: the first when the particle is outside the
spherical boundary and the second when the particle is inside the boundary.
However, the complete treatment of the second case requires the knowledge of
the functions inside the monopole's core, $R_{1l}(r)$ and $R_{2l}(r)$. This
will be done in the next section as applications of the general formalism
developed here, when specific models for the region inside the monopole will
be considered.

\subsection{Particle outside the boundary}

Before considering a specific model for the inner structure of global
monopole, let us investigate the self-energy for the case when the charge is
outside the spherical boundary. Substituting (\ref{goutR}) into (\ref%
{Green-a}), we see that the Green function is expressed in terms of two
contributions:
\begin{equation}
G(\mathbf{x},\mathbf{x}^{\prime })=G_{\mathrm{gm}}(\mathbf{x},\mathbf{x}%
^{\prime })+G_{b}(\mathbf{x},\mathbf{x}^{\prime })\ ,
\end{equation}%
where
\begin{equation}
G_{\mathrm{gm}}(\mathbf{x},\mathbf{x}^{\prime })=\frac{1}{4\pi \alpha ^{2}%
\sqrt{rr^{\prime }}}\sum_{l=0}^{\infty }(2l+1)I_{\nu _{l}}(mr_{<})K_{\nu
_{l}}(mr_{>})P_{l}(\cos \gamma )  \label{G-gm}
\end{equation}%
and
\begin{equation}
G_{b}(\mathbf{x},\mathbf{x}^{\prime })=-\frac{1}{4\pi \alpha ^{2}\sqrt{%
rr^{\prime }}}\sum_{l=0}^{\infty }(2l+1)K_{\nu _{l}}(mr)K_{\nu
_{l}}(mr^{\prime })\frac{\bar{I}_{\nu _{l}}(mR)}{\bar{K}_{\nu _{l}}(mR)}%
P_{l}(\cos \gamma )\ .  \label{Gb0}
\end{equation}%
The first part corresponds to the Green function for the geometry of a
point-like global monopole in the absence of boundary and the second one is
induced by the boundary. In the formulas above, $\gamma $ is the angle
between the directions $(\theta ,\ \varphi )$ and $(\theta ^{\prime },\
\varphi ^{\prime })$ and $P_{l}(x)$ represents the Legendre polynomials of
degree $l$.

The induced scalar self-energy is obtained by taking the coincidence limit
in the renormalized Green function. We shall prove later that for points
with $r>R$, the boundary-induced term, (\ref{Gb0}), is finite and the
divergence takes place in the point-like monopole part only. So, to provide
a finite and well defined result for (\ref{ERen}) we have to renormalize the
Green function $G_{\mathrm{gm}}(\mathbf{x},\mathbf{x}^{\prime })$ only.
First of all we may take $\gamma =0$ in the above expressions. The
renormalized Green function is expressed by:
\begin{equation}
G_{\mathrm{Ren}}(r_{p},r_{p})=G_{\mathrm{gm,ren}%
}(r_{p},r_{p})+G_{b}(r_{p},r_{p})\ ,  \label{Gren}
\end{equation}%
where
\begin{equation}
G_{\mathrm{gm,ren}}(r_{p},r_{p})=\lim_{r\rightarrow r_{p}}[G_{\mathrm{gm}%
}(r,r_{p})-G_{\mathrm{Sing}}(r,r_{p})]\ .
\end{equation}%
In the region outside the monopole's core the radial one-half of geodesic
distance becomes $|r-r^{\prime }|/2$, consequently
\begin{equation}
G_{\mathrm{Sing}}(r,r^{\prime })=\frac{1}{4\pi }\left( \frac{1}{|r-r^{\prime
}|}-m\right) \ .
\end{equation}

Now, by using $G_{\mathrm{gm}}(r,r^{\prime })$ given in (\ref{G-gm}), we
have:
\begin{equation}
G_{\mathrm{gm,ren}}(r_{p},r_{p})=\frac{1}{4\pi r_{p}}\lim_{r^{\prime
}\rightarrow r_{p}}\left[ \frac{1}{\alpha ^{2}}\sum_{l=0}^{\infty
}(2l+1)I_{\nu _{l}}(mr_{<})K_{\nu _{l}}(mr_{>})-\frac{1}{1-t}\right] +\frac{m%
}{4\pi }\ ,  \label{Gmren1}
\end{equation}%
where $t=r_{<}/r_{>}$. In order to evaluate the limit on the right-hand side
of this equation, we note that
\begin{equation}
\lim_{t\rightarrow 1}\left( \frac{1}{\alpha }\sum_{l=0}^{\infty }t^{l/\alpha
+1/2\alpha -1/2}-\frac{1}{1-t}\right) =0\ .  \label{RelLim}
\end{equation}%
So, as a consequence of this relation and replacing in (\ref{Gmren1}) the
expression $1/(1-t)$ by the first term in the brackets in (\ref{RelLim}), we
find\footnote{%
The expression for the renormalized Green function has also been obtained in
\cite{Barbosa2}.}
\begin{equation}
G_{\mathrm{gm,ren}}(r_{p},r_{p})=\frac{S_{(\alpha )}(mr_{p})}{4\pi \alpha
r_{p}}+\frac{m}{4\pi }\ ,  \label{Gmren2}
\end{equation}%
with
\begin{equation}
S_{(\alpha )}(x)=\sum_{l=0}^{\infty }\left[ \frac{2l+1}{\alpha }I_{\nu
_{l}}(x)K_{\nu _{l}}(x)-1\right] \ .  \label{S}
\end{equation}%
Finally the complete expression for the scalar self-energy reads
\begin{equation}
E_{\mathrm{Ren}}=-\frac{q^{2}}{8\pi \alpha r_{p}}S_{(\alpha )}(mr_{p})-\frac{%
q^{2}m}{8\pi }+\frac{q^{2}}{8\pi \alpha ^{2}r_{p}}\sum_{l=0}^{\infty
}(2l+1)K_{\nu _{l}}^{2}(mr_{p})\frac{\bar{I}_{\nu _{l}}(mR)}{\bar{K}_{\nu
_{l}}(mR)}\ .  \label{TE}
\end{equation}

The convergence of the boundary-induced part on the scalar self-energy can
be investigated by analyzing the general term inside the summation of (\ref%
{TE}) for large values of quantum number $l$, using the uniform asymptotic
expansions for large orders of the modified Bessel functions \cite{Abra}:
\begin{eqnarray}
I_{\nu }(\nu z) &\sim &\frac{1}{\sqrt{2\pi \nu }}\frac{e^{\nu \eta (z)}}{%
(1+z^{2})^{1/4}}\sum_{k=0}^{\infty }\frac{u_{k}(w(z))}{\nu ^{k}}\ ,  \notag
\\
\ K_{\nu }(\nu z) &\sim &\sqrt{\frac{\pi }{2\nu }}\frac{e^{-\nu \eta (z)}}{%
(1+z^{2})^{1/4}}\sum_{k=0}^{\infty }(-1)^{k}\frac{u_{k}(w(z))}{\nu ^{k}}\ ,
\label{UAE}
\end{eqnarray}%
where%
\begin{equation}
w(z)=\frac{1}{\sqrt{1+z^{2}}},\;\eta (z)=\sqrt{1+z^{2}}+\ln \left( \frac{z}{%
1+\sqrt{1+z^{2}}}\right) .  \label{etaw}
\end{equation}%
The relevant expressions for $u_{k}(z)$ can be found in \cite{Abra}. In our
expression $z=mr/\nu $. Keeping only the leading terms in $1/\nu _{l}$ for
the above expansions, we find
\begin{eqnarray}
K_{\nu _{l}}(mr) &\approx &\sqrt{\frac{\pi }{2\nu _{l}}}\ \left( \frac{2\nu
_{l}}{emr}\right) ^{\nu _{l}}\ ,  \notag  \label{asymptKI} \\
I_{\nu _{l}}(mr) &\approx &\frac{1}{\sqrt{2\pi \nu _{l}}}\ \left( \frac{emr}{%
2\nu _{l}}\right) ^{\nu _{l}}\ .
\end{eqnarray}%
Taking the above expressions and the fact that $2\alpha \nu _{l}\approx 2l+1+%
\frac{(1-\alpha ^{2})(8\xi -1)}{2(2l+1)}$, after some intermediate steps we
find that for large $l$ the term inside the summation of \eqref{TE} behaves
as:
\begin{equation}
\pm \alpha \left( R/r_{p}\right) ^{2l/\alpha }\ ,
\end{equation}%
and the series converges for $r_{p}>R$. The positive (negative) sign in the
above result corresponds to DBC (NBC).

At large distances from the boundary, assuming that $mr_{p}\gg 1$, it can be
seen that the boundary-induced part in (\ref{TE}) is suppressed by the
factor $e^{-2mr_{p}}$ and the boundary-free part dominates. With respect the
dependence of the curvature coupling parameter the corresponding force can
be either attractive or repulsive (see the numerical examples below). Near
the boundary the renormalized self-energy is dominated by the
boundary-induced term and behaves as,
\begin{equation}
E_{\mathrm{Ren}}\approx \pm \frac{q^{2}}{16\pi \alpha }\frac{R^{(1-\alpha
)/\alpha }}{r_{p}^{1/\alpha }-R^{1/\alpha }}\ .  \label{E-out}
\end{equation}%
The upper (lower) sign corresponds to DBC (NBC). Note that the leading term
does not depend on the mass and on the curvature coupling parameter. From (%
\ref{E-out}) we conclude that near the spherical boundary the self-force on
a charged particle is repulsive with respect to the boundary for DBC and
attractive for NBC.

The expression for the self-energy is simplified in the case of a massless
field. By using the formulas for the modified Bessel functions for small
arguments, from general formula (\ref{TE}) one finds%
\begin{equation}
E_{\mathrm{Ren}}=-\frac{q^{2}}{8\pi \alpha r_{p}}\left[ S_{(\alpha
)}(0)-\sum_{l=0}^{\infty }\frac{2l+1}{2\alpha \nu _{l}}F_{l}\left( \frac{R}{%
r_{p}}\right) ^{2\nu _{l}}\right] \ .  \label{TEm0}
\end{equation}%
In this formula%
\begin{equation}
S_{(\alpha )}(0)=\sum_{l=0}^{\infty }\left( \frac{2l+1}{2\alpha \nu _{l}}%
-1\right) ,  \label{S0}
\end{equation}%
and we have defined%
\begin{equation}
F_{l}=\left\{
\begin{array}{ll}
1, & \text{for DBC} \\
-\left( 2\nu _{l}-1\right) /\left( 2\nu _{l}+1\right) , & \text{for NBC}%
\end{array}%
\right. .  \label{Fl}
\end{equation}%
Note that $S_{(\alpha )}(0)>0$ for $\xi <1/8$ and $S_{(\alpha )}(0)<0$ for $%
\xi >1/8$. For the special case $\xi =1/8$ one has $S_{(\alpha )}(0)=0$. At
large distances from the sphere, $r_{p}\gg R$, the boundary-induced part is
suppressed with respect to the boundary-free part by the factor $%
(R/r_{p})^{2\nu _{0}}$. In this region the self-force for a massless field
is attractive for $\xi <1/8$ and repulsive for $\xi >1/8$. In the absence of
the global monopole, taking $\alpha =1$, from (\ref{TEm0}) for the case of
DBC we get:%
\begin{equation}
E_{\mathrm{Ren}}^{(D)}|_{\alpha =1}\ =\frac{q^{2}}{8\pi }\frac{R}{%
r_{p}^{2}-R^{2}}.  \label{EDalf1}
\end{equation}%
The corresponding force is repulsive. The expression of the energy for NBC
takes the form%
\begin{equation}
E_{\mathrm{Ren}}^{(N)}|_{\alpha =1}\ =-E_{\mathrm{Ren}}^{(D)}|_{\alpha =1}\ -%
\frac{q^{2}}{8\pi R}\ln (1-R^{2}/r_{p}^{2}),  \label{ENalf1}
\end{equation}%
and the force is attractive. Note that at large distances from the sphere
the decay of the energy in the case of NBC is stronger: $E_{\mathrm{Ren}%
}^{(N)}|_{\alpha =1}\propto 1/r_{p}^{4}$. Taking the limit $%
R,r_{p}\rightarrow \infty $ with $r_{p}-R$ fixed, we obtain from (\ref%
{EDalf1}) and (\ref{ENalf1}) the interaction energies for a scalar charge
with a flat boundary. The corresponding expressions are easily obtained by
using the image method. For DBC the sign of the image charge is opposite and
the corresponding force is repulsive (for the interaction of scalar
point-charges see \cite{Ande67}). For the NBC the normal derivative of the
field is zero on the flat boundary and the image charge has the same sign
with the attractive force.

\begin{figure}[tbph]
\begin{center}
\begin{tabular}{cc}
\epsfig{figure=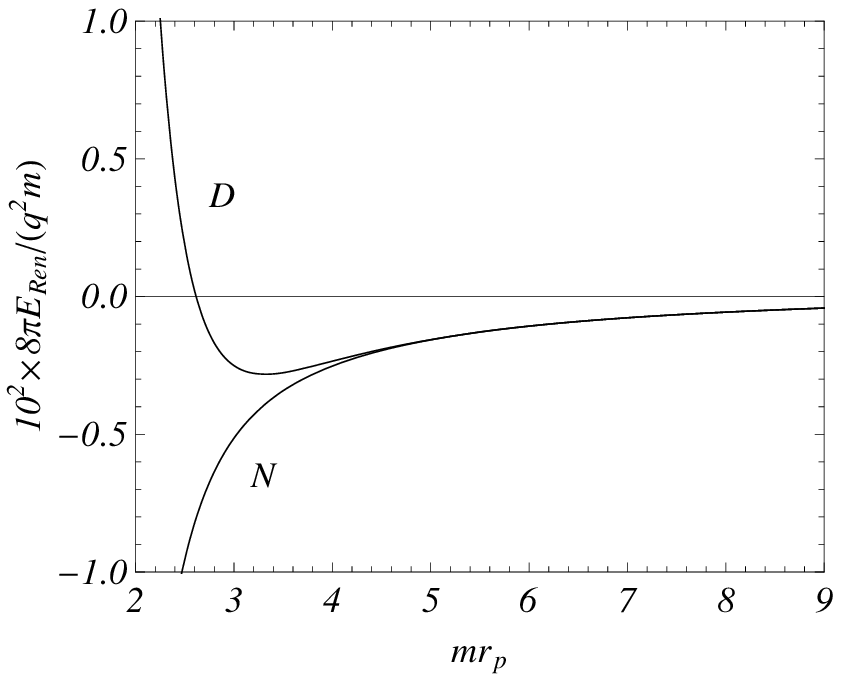, width=7.cm, height=6.5cm,angle=0} & \quad %
\epsfig{figure=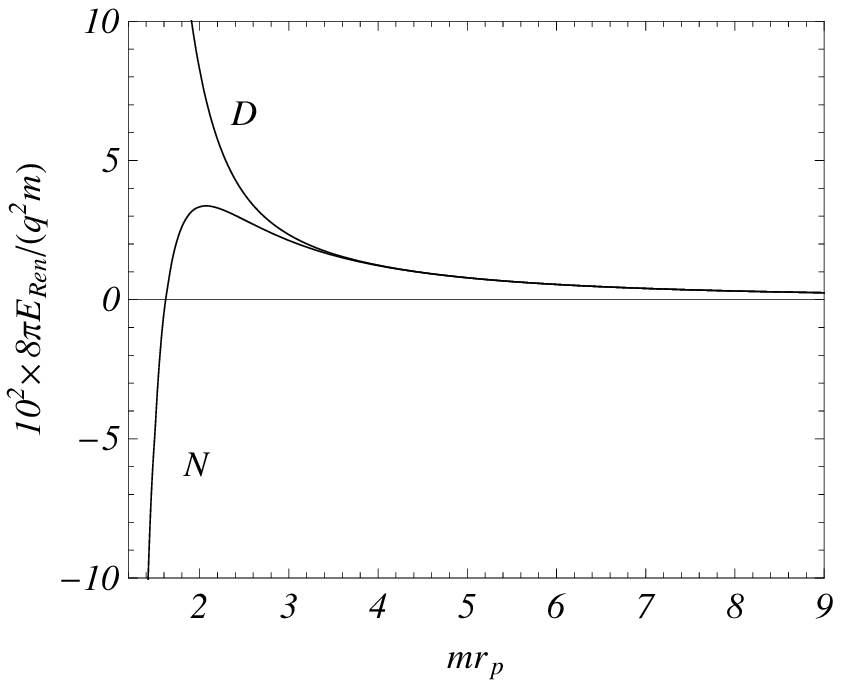, width=7.cm, height=6.5cm,angle=0}%
\end{tabular}%
\end{center}
\caption{These graphs provide the behavior of the self-energy, $10^{2}\times
8\protect\pi E_{\mathrm{Ren}}/(q^{2}m)$, as a function of $mr_{p}$
considering $\protect\xi =0$ in the left panel and $\protect\xi =1$ in the
right panel. For both cases we adopted $\protect\alpha =0.9$ and $mR=1$. The
capital letters near the curves represent the boundary condition obeyed by
the field, named Dirichlet (D) and Neumann (N), ones.}
\label{fig1}
\end{figure}

In figure \ref{fig1} we exhibit the behavior for the self-energy given by (%
\ref{TE}) as a function of $mr_{p}$ for fields obeying DBC and NBC. Here and
in the figures below the capital letters near the curves, D and N, represent
the respective boundary conditions. In the left panel we consider $\xi =0$
and for the right panel $\xi =1$. In both cases we adopted $\alpha =0.9$ and
$mR=1$. By these plots we can conclude that the boundary-induced
contributions do not depend strongly with the curvature coupling, $\xi $,
and for points near the boundary, i.e., with $mr_{p}$ near the unity, these
contributions dominate the behavior of the self-energy; on the other hand
for points far from the boundary, the boundary-free contributions dominate;
the latter are negative for $\xi =0$ and positive for $\xi =1$. At large
distances from the boundary, the self-force is attractive with respect to
the boundary in the first case and repulsive in the second case. Finally we
also can see that for $\xi =0$, the self-energy presents a stable
equilibrium point when the field obeys DBC.

\subsection{Particle inside the boundary}

The second analysis is for the case when the particle is inside the
spherical boundary. Two distinct situations take place: For the particle
outside the monopole's core, $r^{\prime }\geqslant a$, and for the particle
inside the core, $r^{\prime }\leqslant a$. These two situations will be
considered separately in the following.

For the particle outside the monopole, the corresponding Green function can
be obtained by substituting \eqref{gout+} into (\ref{Green-a}). This
procedure allows us to write,
\begin{equation}
G(\mathbf{x},\mathbf{x}^{\prime })=G_{\mathrm{gm}}(\mathbf{x},\mathbf{x}%
^{\prime })+G_{c}(\mathbf{x},\mathbf{x}^{\prime })+G_{b}(\mathbf{x},\mathbf{x%
}^{\prime })\ ,
\end{equation}%
where $G_{\mathrm{gm}}(\mathbf{x},\mathbf{x}^{\prime })$ is given by %
\eqref{G-gm},
\begin{equation}
G_{c}(\mathbf{x},\mathbf{x}^{\prime })=\frac{1}{4\pi \alpha ^{2}\sqrt{%
rr^{\prime }}}\sum_{l=0}^{\infty }(2l+1)D_{l}^{(+)}(a)K_{\nu
_{l}}(mr^{\prime })K_{\nu _{l}}(mr)P_{l}(\cos \gamma )  \label{Gc}
\end{equation}%
is the contribution induced by the non-trivial structure of the monopole
when the boundary is absent and
\begin{equation}
G_{b}(\mathbf{x},\mathbf{x}^{\prime })=-\frac{1}{4\pi \alpha ^{2}\sqrt{%
rr^{\prime }}}\sum_{l=0}^{\infty }(2l+1)M_{\nu _{l}}(mr)M_{\nu
_{l}}(mr^{\prime })\frac{\bar{K}_{\nu _{l}}(mR)}{\bar{M}_{\nu _{l}}(mR)}%
P_{l}(\cos \gamma )  \label{Gb}
\end{equation}%
is the contribution induced by the boundary. We can see that (\ref{Gc}) and (%
\ref{Gb}) contain information about the structure of the monopole' core,
through the coefficient $D_{l}^{(+)}(a)$ given by \eqref{D+}.

The induced scalar self-energy is obtained by taking the coincidence limit
in the renormalized Green function. As it will be shown later, the only
divergence that appears in this limit is in the point-like monopole part.
So, in order to provide a finite and well defined result for (\ref{ERen}),
we have to renormalize the Green function $G_{\mathrm{gm}}(\mathbf{x},%
\mathbf{x}^{\prime })$. In fact this has already been done in the last
subsection and the result is given by \eqref{Gmren2}. Finally the complete
expression for scalar self-energy reads
\begin{eqnarray}
E_{\mathrm{Ren}} &=&-\frac{q^{2}}{8\pi \alpha r_{p}}S_{(\alpha )}(mr_{p})-%
\frac{q^{2}m}{8\pi }-\frac{q^{2}}{8\pi \alpha ^{2}r_{p}}\sum_{l=0}^{\infty
}(2l+1)D_{l}^{(+)}(a)K_{\nu _{l}}^{2}(mr_{p})  \notag \\
&+&\frac{q^{2}}{8\pi \alpha ^{2}r_{p}}\sum_{l=0}^{\infty }(2l+1)M_{\nu
_{l}}^{2}(mr_{p})\frac{\bar{K}_{\nu _{l}}(mR)}{\bar{M}_{\nu _{l}}(mR)}\ .
\label{TE1}
\end{eqnarray}%
In \cite{Barbosa2}, we have shown that the core-induced part in the
self-energy above is finite for points away from the core. It remains to
analyze the boundary-induced part for point near the boundary itself.

Before to enter in the details about the boundary-induced part considering
specific models for the monopole core, let us first investigate this term
for the simpler case considering the monopole as a point-like defect. For
this case we have to take the limit $a\rightarrow 0$ in (\ref{TE1}). By
analyzing the behavior of the modified Bessel functions for small arguments,
we can verify that $D_{l}^{(+)}(a)$, given in \eqref{D+}, goes to zero. This
implies that $M_{\nu }(z)\rightarrow I_{\nu }(z)$, consequently the
self-energy reads,
\begin{equation}
E_{\mathrm{Ren}}=-\frac{q^{2}}{8\pi \alpha r_{p}}S_{(\alpha )}(mr_{p})-\frac{%
q^{2}m}{8\pi }+\frac{q^{2}}{8\pi \alpha ^{2}r_{p}}\sum_{l=0}^{\infty
}(2l+1)I_{\nu _{l}}^{2}(mr_{p})\frac{\bar{K}_{\nu _{l}}(mR)}{\bar{I}_{\nu
_{l}}(mR)}\ .  \label{TE2}
\end{equation}%
For $\xi >0$ the boundary-induced part in (\ref{TE2}) vanishes for $r_{p}=0$%
. For a minimally coupled field, $\xi =0$, the boundary-induced part is
finite at $r_{p}=0$ with the only non-zero contribution coming from the $l=0$
term. Consequently, near the global monopole, $mr_{p}\ll 1$, the
boundary-free part dominates and to the leading order one has $E_{\mathrm{Ren%
}}\approx -q^{2}S_{(\alpha )}(0)/(8\pi \alpha r_{p})$. In this region the
self-force is attractive with respect to the global monopole for $\xi <1/8$
and repulsive for $\xi >1/8$.

For a massless field, the general expression (\ref{TE2}) takes the form%
\begin{equation}
E_{\mathrm{Ren}}=-\frac{q^{2}}{8\pi \alpha r_{p}}\left[ S_{(\alpha
)}(0)-\sum_{l=0}^{\infty }\frac{2l+1}{2\alpha \nu _{l}F_{l}}\left( \frac{%
r_{p}}{R}\right) ^{2\nu _{l}}\right] \ ,  \label{TE2m0}
\end{equation}%
where $F_{l}$ is given by (\ref{Fl}). In deriving (\ref{TE2m0}) for NBC we
have assumed that $\nu _{l}\neq 1/2$. If for some $l=l_{0}$ one has $\nu
_{l_{0}}=1/2$, the expression for the self-energy is given by (\ref{TE2m0})
where now the term $l=l_{0}$ must be excluded from the summation and the
corresponding contribution should be added separately. The latter does not
depend on $r_{p}$ and has the form $q^{2}(2l_{0}+1)/(8\pi \alpha ^{2}R)$. In
the absence of the global monopole ($\alpha =1$), from (\ref{TE2m0}) the
following results are obtained
\begin{eqnarray}
E_{\mathrm{Ren}}^{(D)}|_{\alpha =1} &=&\frac{q^{2}}{8\pi }\frac{R}{%
R^{2}-r_{p}^{2}},  \notag \\
E_{\mathrm{Ren}}^{(N)}|_{\alpha =1} &=&-E_{\mathrm{Ren}}^{(D)}|_{\alpha =1}\
+\frac{q^{2}}{8\pi R}\ln (1-r_{p}^{2}/R^{2}).  \label{EDNalf1}
\end{eqnarray}%
As for the exterior region, the corresponding force is repulsive/attractive
with respect to the sphere for DBC/NBC.

We can verify the convergence of the boundary-induced part, by analyzing the
term inside the summation for large values of $l$. Developing a similar
procedure as we did in the last section for the core-induced part, we arrive
at the following result:
\begin{equation}
\pm \alpha \left( r_{p}/R\right) ^{2l/\alpha }\ ,
\end{equation}%
with positive (negative) sign for DBC (NBC). The corresponding series
converge for $r_p<R$.

Near the boundary the renormalized scalar-energy is dominated by the
boundary-induced term and it behaves as,
\begin{equation}
E_{\mathrm{Ren}}\approx \pm \frac{q^{2}}{16\pi \alpha }\frac{R^{(1-\alpha
)/\alpha }}{R^{1/\alpha }-r_{p}^{1/\alpha }}\ .  \label{ER+}
\end{equation}%
Here also, the positive (negative) sign corresponds to DBC (NBC). In this
region the self-force is repulsive with respect to the boundary for DBC and
attractive for NBC.

\begin{figure}[tbph]
\begin{center}
\begin{tabular}{cc}
\epsfig{figure=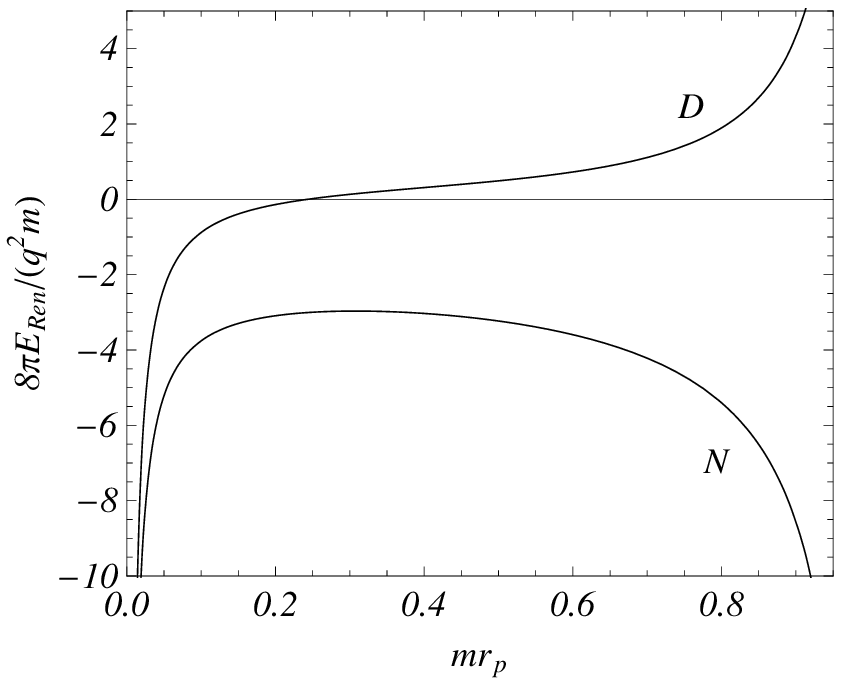, width=7.cm, height=6.5cm,angle=0} & \quad %
\epsfig{figure=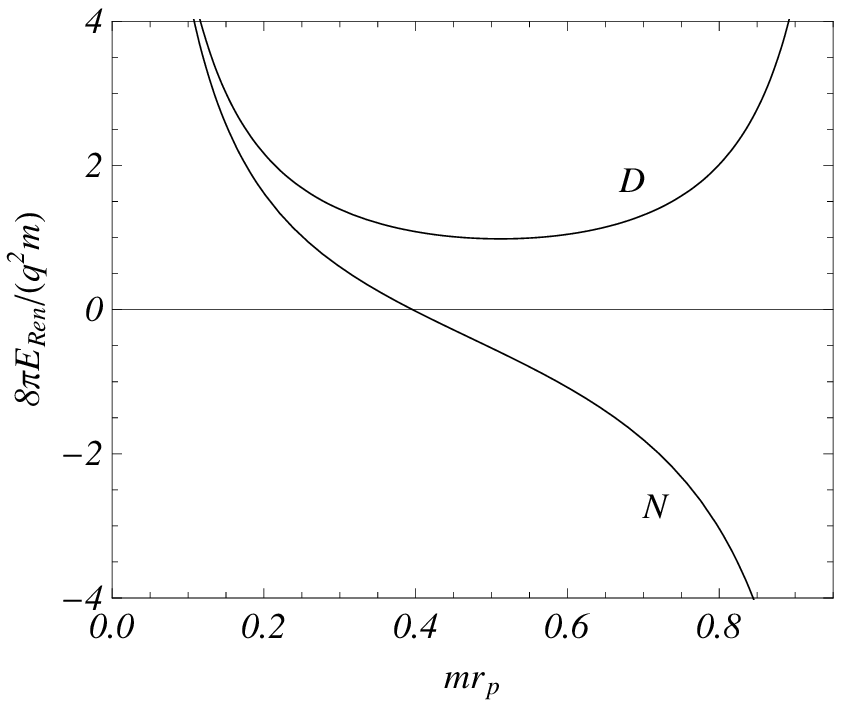, width=7.cm, height=6.5cm,angle=0}%
\end{tabular}%
\end{center}
\caption{The self-energy inside a spherical boundary as a function of $%
mr_{p} $ considering $\protect\xi =0$ in the left panel and $\protect\xi =1$
in the right panel. For both cases we adopted $\protect\alpha =0.9$ and $%
mR=1 $. }
\label{fig2}
\end{figure}

In figure \ref{fig2} we have plotted the self-energy given by (\ref{TE2})
versus $mr_{p}$ for fields with DBC and NBC. The left and right panels are
for $\xi =0$ and $\xi =1$, respectively. In both cases the graphs are
plotted for $\alpha =0.9$ and $mR=1$. Also here, we can conclude that the
boundary-induced contributions do not depend strongly with the curvature
coupling, $\xi $, for points near the boundary; in fact these contributions
dominate the behavior of the self-energy in this region. On the other hand
for points near the core's center the boundary-free contributions dominate;
the latter are negative for $\xi =0$ and positive for $\xi =1$. Finally we
observe that for $\xi =1$, the self-energy presents a stable equilibrium
point for the field with DBC.

Considering now the general situation, we can see that for the particle
inside the core the corresponding Green function can be formally expressed
by substituting \eqref{gin-} into \eqref{Green-a}:
\begin{equation}
G(\mathbf{x},\mathbf{x}^{\prime })=G_{0}(\mathbf{x},\mathbf{x}^{\prime
})+G_{\alpha ,c}(\mathbf{x},\mathbf{x}^{\prime })\ ,
\end{equation}%
where
\begin{equation}
G_{0}(\mathbf{x},\mathbf{x}^{\prime })=\frac{1}{4\pi }\sum_{l=0}^{\infty
}(2l+1)R_{1l}(r_{<})R_{2l}(r_{>})P_{l}(\cos \gamma )\ ,  \label{G0in}
\end{equation}%
is the Green function for the background geometry described by the line
element (\ref{gm1}) and the term
\begin{equation}
G_{\alpha ,b}(\mathbf{x},\mathbf{x}^{\prime })=\frac{1}{4\pi }%
\sum_{l=0}^{\infty }(2l+1)D_{l}^{(-)}(a)R_{1l}(r)R_{1l}(r^{\prime
})P_{l}(\cos \gamma )\ ,  \label{Galfain}
\end{equation}%
is due to the global monopole geometry in the region $r>a$ and the boundary
condition obeyed by the field at $r=R$. This information is explicitly
contained in the coefficient $D_{l}^{(-)}(a)$ given by \eqref{D-}. We can
see that for $mR\rightarrow \infty $, this coefficient reduces to
\begin{equation}
D_{l}^{(-)}(a)=-\frac{a{\mathcal{R}}_{l}^{(2)}(a)K_{\nu _{l}}(ma)-R_{2l}(a)%
\tilde{K}_{\nu _{l}}(ma)}{a{\mathcal{R}}_{l}^{(1)}(a)K_{\nu
_{l}}(ma)-R_{1l}(a)\tilde{K}_{\nu _{l}}(ma)}\ .
\end{equation}%
For points away from the core's boundary, \eqref{Galfain} is finite in the
coincidence limit.

The renormalized scalar self-energy for a charge inside the core is written
in the form
\begin{equation}
E_{\mathrm{Ren}}=-\frac{q^{2}}{2}G_{0,\mathrm{Ren}}(\mathbf{x}_{p},\mathbf{x}%
_{p})-\frac{q^{2}}{8\pi }\sum_{l=0}^{\infty
}(2l+1)D_{l}^{(-)}(a)R_{1l}^{2}(r_{p})\ ,  \label{ERin}
\end{equation}%
where the renormalized Green function is given by
\begin{equation}
G_{0,\mathrm{Ren}}(\mathbf{x}_{p},\mathbf{x}_{p})=\lim_{\mathbf{x}%
\rightarrow \mathbf{x}_{p}}\left[ G_{0}(\mathbf{x},\mathbf{x}_{p})-G_{%
\mathrm{Sing}}(\mathbf{x},\mathbf{x}_{p})\right] \ .  \label{G0renin}
\end{equation}%
Because the divergent part of the Green function should have the same
structure as (\ref{Had3D}), the above expression provides a finite result.

\section{Applications}

\label{Applic}

As we have mentioned in the Introduction, there is no closed expression for
the metric tensor in the region inside the global monopole. This situation
also happens for the cosmic string spacetime; however two different models
have been proposed to describe the geometry inside the string's core, named
the ballpoint pen and flower-pot models proposed in \cite{Gott} and \cite{BA}%
, respectively. These two models have been also used in \cite{Barbosa2} to
analyze the self-energy associated with a massive scalar field in the global
monopole spacetime. So, motivated by this result, we decided to continue
along the same line of investigation considering here both models, and
analyzing the influence on the self-energies induced by the spherical
boundary.

\subsection{Flower-pot model}

\label{Flower}

The first model to be considered is the flower-pot one. For this model the
interior line element has the form
\begin{equation}
ds^{2}=-dt^{2}+dr^{2}+\left[ r+(\alpha -1)a\right] ^{2}(d^{2}\theta +\sin
^{2}\theta d^{2}\varphi )\ .  \label{intflow}
\end{equation}%
This spacetime corresponds to Minkowski one which can be easily verified by
expressing the above line element in terms of the new radial coordinate, $%
\tilde{r}=r+(\alpha -1)a$. As we have mentioned before, from the Israel
matching conditions for the metric tensors corresponding to (\ref{gm}) and (%
\ref{intflow}), we find the singular contribution for the Ricci scalar
curvature located on the core's surface $r=a$ \cite{Mello5}:
\begin{equation}
\bar{R}=4\frac{1-\alpha }{\alpha a}\ .  \label{Rbar}
\end{equation}

In the region inside the monopole's core the two linearly independent
solutions for the radial functions are:
\begin{equation}
R_{1l}(r)=\frac{I_{l+1/2}(m\tilde{r})}{\sqrt{\tilde{r}}}\ \mathrm{and}\
R_{2l}(r)=\frac{K_{l+1/2}(m\tilde{r})}{\sqrt{\tilde{r}}}\ .
\label{Flower-Int}
\end{equation}%
With these two solutions, we can write explicitly the Green functions in
both, interior and exterior regions, and consequently the corresponding
self-energies. These expressions depend on the coefficients $D_{l}^{(+)}(a)$
and $D_{l}^{(-)}(a)$, which can be explicitly provided.

Considering first the case with the particle outside the core we have,
\begin{equation}
D_{l}^{(+)}(a)=-\frac{n_{l}(a)}{d_{l}(a)}\ ,  \label{D+-}
\end{equation}%
where
\begin{eqnarray}
n_{l}(a) &=&I_{\nu _{l}}(ma)I_{l+1/2}(m\alpha a)\left[ \frac{l+4\xi
(1-\alpha )}{\alpha }-\nu _{l}+\frac{1}{2}\right]  \notag \\
&+&ma\left[ I_{\nu _{l}}(ma)I_{l+3/2}(m\alpha a)-I_{l+1/2}(m\alpha a)I_{\nu
_{l}+1}(ma)\right] \ ,  \label{n+} \\
d_{l}(a) &=&K_{\nu _{l}}(ma)I_{l+1/2}(m\alpha a)\left[ \frac{l+4\xi
(1-\alpha )}{\alpha }-\nu _{l}+\frac{1}{2}\right]  \notag \\
&+&ma\left[ K_{\nu _{l}}(ma)I_{l+3/2}(m\alpha a)+I_{l+1/2}(m\alpha a)K_{\nu
_{l}+1}(ma)\right] \ .  \label{d}
\end{eqnarray}

With this coefficient we are in position to write the expression for the
complete self-energy \eqref{TE1}. By this expression we can investigate the
behavior of the self-energy due to the core-induced part for points near the
core. In fact this analysis has been done in \cite{Barbosa2}, and there we
have proved that the boundary-free part
\begin{equation}
E^{c}=-\frac{q^{2}}{8\pi \alpha ^{2}r_{p}}\sum_{l=0}^{\infty
}(2l+1)D_{l}^{(+)}(a)K_{\nu _{l}}^{2}(mr_{p})\ ,
\end{equation}%
is divergent near the core. In order to verify this fact, we have analyzed
the general term inside the summation for large value of $l$. By making use
of uniform asymptotic expansions (\ref{UAE}), we get
\begin{equation}
d_{l}\left( a/r_{p}\right) ^{2\nu _{l}}\ ,  \label{Approx1}
\end{equation}%
where%
\begin{equation}
d_{l}=\frac{\alpha -2\alpha \nu _{l}+2l+8\xi (1-\alpha )}{\alpha +2\alpha
\nu _{l}+2l+8\xi (1-\alpha )}.  \label{dl}
\end{equation}%
For large value of $l$ we have $2\alpha \nu _{l}\approx (2l+1)+\frac{%
(1-\alpha ^{2})(8\xi -1)}{2(2l+1)}+\cdots $. Substituting this expansion
into (\ref{Approx1}), for the leading term in $1/l$, we obtain
\begin{equation}
\alpha (1-\alpha )\frac{1-8\xi }{4l}\left( \frac{a}{r_{p}}\right)
^{2l/\alpha }\ .
\end{equation}%
Finally, after some intermediate steps we find:
\begin{equation}
E_{\mathrm{Ren}}\approx q^{2}\frac{(1-\alpha )(1-8\xi )}{32\pi \alpha a}\ln %
\left[ 1-\left( a/r_{p}\right) ^{1/\alpha }\right] \ .  \label{FP+}
\end{equation}%
The corresponding self-force is attractive (repulsive) with respect to the
core boundary for $\xi <1/8$ ($\xi >1/8$).

We may also want to investigate the behavior of the boundary-induced term
near the boundary, taking into account this specific model for the core. The
boundary-induced term in \eqref{TE1} is given by,
\begin{equation}
E^{b}=\frac{q^{2}}{8\pi \alpha ^{2}r_{p}}\sum_{l=0}^{\infty }(2l+1)M_{\nu
_{l}}^{2}(mr_{p})\frac{\bar{K}_{\nu _{l}}(mR)}{\bar{M}_{\nu _{l}}(mR)}\ .
\label{Ebb}
\end{equation}%
Remembering that $M_{\nu }(z)=I_{\nu }(z)+D_{l}^{(+)}(a)K_{\nu }(z)$, we can
see that for large values of $l$ the coefficient $D_{l}^{(+)}(a)$ behaves as
shown below:
\begin{equation}
D_{l}^{(+)}(a)\approx \frac{d_{l}}{\alpha \pi }\ \left( \frac{ema}{2\nu _{l}}%
\right) ^{2\nu _{l}}.
\end{equation}%
Adopting also for the modified Bessel functions their asymptotic expansions,
we may verify that,
\begin{equation}
M_{\nu _{l}}(z)\approx \frac{1}{\sqrt{2\pi \nu _{l}}}\left( \frac{ez}{2\nu
_{l}}\right) ^{\nu _{l}}\left[ 1+\frac{d_{l}}{\alpha }\left( \frac{ma}{z}%
\right) ^{2\nu _{l}}\right] \approx I_{\nu _{l}}(z)
\end{equation}%
for $z=mR$ or $mr_{p}$. So, near the boundary, the dominant contribution on %
\eqref{Ebb} is independent of the core, consequently, $E^{b}$ can be given
by \eqref{ER+}.

In the special case of a massless scalar field, the general expression for
the self-energy in the region $a\leqslant r_{p}\leqslant R$ reduces to%
\begin{equation}
E_{\mathrm{Ren}}=-\frac{q^{2}}{8\pi \alpha r_{p}}S_{(\alpha )}(0)+\frac{q^{2}%
}{8\pi r_{p}}\sum_{l=0}^{\infty }\frac{2l+1}{2\alpha ^{2}\nu _{l}}\left\{
d_{l}+\frac{\left[ (r_{p}/a)^{2\nu _{l}}-d_{l}\right] ^{2}}{F_{l}(R/a)^{2\nu
_{l}}-d_{l}}\right\} \left( \frac{a}{r_{p}}\right) ^{2\nu _{l}},
\label{Efpm0}
\end{equation}%
with $F_{l}$ and $d_{l}$ defined by the relations (\ref{Fl}) and (\ref{dl}).

Now let us turn our investigation of the self-energy for the region inside
the monopole's core. Substituting the functions (\ref{Flower-Int}) into the
formulas (\ref{G0in}) and (\ref{Galfain}) for the corresponding Green
function in the interior region one finds \cite{Abra},
\begin{equation}
G_{0}(\mathbf{x},\mathbf{x}^{\prime })=\frac{1}{4\pi }\frac{e^{-m\tilde{R}}}{%
\tilde{R}}\ ,  \label{G00}
\end{equation}%
with $\tilde{R}=\sqrt{(\tilde{r}^{\prime })^{2}+(\tilde{r})^{2}-2\tilde{r}%
\tilde{r}^{\prime }\cos \gamma }$, being $\gamma $ the angle between the two
directions defined by the unit vectors $\hat{\tilde{r}}^{\prime }$ and $\hat{%
\tilde{r}}$. Taking $\gamma =0$ we get $\tilde{R}=|r-r^{\prime }|$. Because
in the flower-pot model the geometry in the region inside the core is
Minkowski one, we have $G_{0}(\mathbf{x},\mathbf{x}^{\prime })=G_{\mathrm{%
Sing}}(\mathbf{x},\mathbf{x}^{\prime })$, consequently $G_{0,\mathrm{Ren}}(%
\mathbf{x}_{p},\mathbf{x}_{p})=0$. Consequently, the scalar self-energy in
this region is given by
\begin{equation}
E_{\mathrm{Ren}}=-\frac{q^{2}}{8\pi \tilde{r}_{p}}\sum_{l=0}^{\infty
}(2l+1)D_{l}^{(-)}(a)I_{l+1/2}^{2}(m\tilde{r}_{p})\ ,  \label{E-ind}
\end{equation}%
being $\tilde{r}_{p}=r_{p}+(\alpha -1)a$. Due to the fact that the
expression derived from \eqref{D-} is too long, we shall not reproduce it in
its complete form, only its asymptotic behavior for specific values of $l$
will be provided. Near the core's center, $m\tilde{r}_{p}\ll 1$,
\begin{equation}
I_{l+1/2}(m\tilde{r}_{p})\approx \left( \frac{m\tilde{r}_{p}}{2}\right)
^{l+1/2}\frac{1}{\Gamma (l+3/2)}\ ,
\end{equation}%
so the main contribution into the self-energy comes from the lowest mode, $%
l=0$, resulting in
\begin{equation}
E_{\mathrm{Ren}}\approx -\frac{q^{2}mD_{0}^{(-)}(a)}{4\pi ^{2}}\ .
\label{EE}
\end{equation}%
We can see that for $R\gg a$ the leading term in $D_{0}^{(-)}(a)$ is
boundary independent.

Also we can evaluate the scalar self-energy near the core's boundary.
Because $D_{l}^{(-)}(a)$ presents contributions due to the boundary, to
analyze possible divergent contribution of \eqref{E-ind}, we can use
asymptotic expansions for large order of the modified Bessel functions.
Doing this, after many steps, we find that the numerator, $n_{l}(a)$, and
denominator, $d_{l}(a)$, of \eqref{D-} are dominated by the terms involving $%
\bar{I}_{\nu _{l}}(mR)$. Consequently the leading terms of $D_{0}^{(-)}(a)$,
coincide with the corresponding coefficient when we take $R\rightarrow
\infty $, providing for the self-energy the same singular behavior as for
the boundary-free geometry. The leading term inside the summation in (\ref%
{E-ind}) behaves as,
\begin{equation}
\frac{(1-\alpha )(1-8\xi )}{4l\alpha a}\left( \frac{\tilde{r}}{\alpha a}%
\right) ^{2l}\ .
\end{equation}%
Finally, taking this expression back into (\ref{E-ind}) we obtain,
\begin{equation}
E_{\mathrm{Ren}}\approx q^{2}\frac{(1-\alpha )(1-8\xi )}{32\pi \alpha a}\ln %
\left[ 1-\frac{\tilde{r}_{p}}{\alpha a}\right] \ .  \label{FP-}
\end{equation}%
This term does not depend on the mass. As it is seen from (\ref{FP-}), near
the core's boundary the self-force is attractive with respect to this
boundary for $\xi <1/8$ and repulsive for $\xi >1/8$.

For a massless scalar field, the expression for the self energy inside the
core of the monopole is obtained from (\ref{E-ind}) by using the formulas
for the modified Bessel functions for small arguments:%
\begin{equation}
E_{\mathrm{Ren}}=\frac{q^{2}}{8\pi \alpha a}\sum_{l=0}^{\infty }\frac{%
(b_{l}^{(+)}-l-1)F_{l}-(b_{l}^{(-)}-l-1)(a/R)^{2\nu _{l}}}{%
(b_{l}^{(+)}+l)F_{l}-(b_{l}^{(-)}+l)(a/R)^{2\nu _{l}}}\left( \frac{\tilde{r}%
_{p}}{\alpha a}\right) ^{2l}\ ,  \label{Efpinm0}
\end{equation}%
where $F_{l}$ is defined by (\ref{Fl}) and%
\begin{equation}
b_{l}^{(\pm )}=4\xi (1-\alpha )+\alpha /2\pm \alpha \nu _{l}.  \label{blpm}
\end{equation}%
From this formula the asymptotic behavior (\ref{FP-}) is easily obtained for
points near the core boundary. The corresponding force linearly vanishes at
the core center.

\begin{figure}[tbph]
\begin{center}
\begin{tabular}{cc}
\epsfig{figure=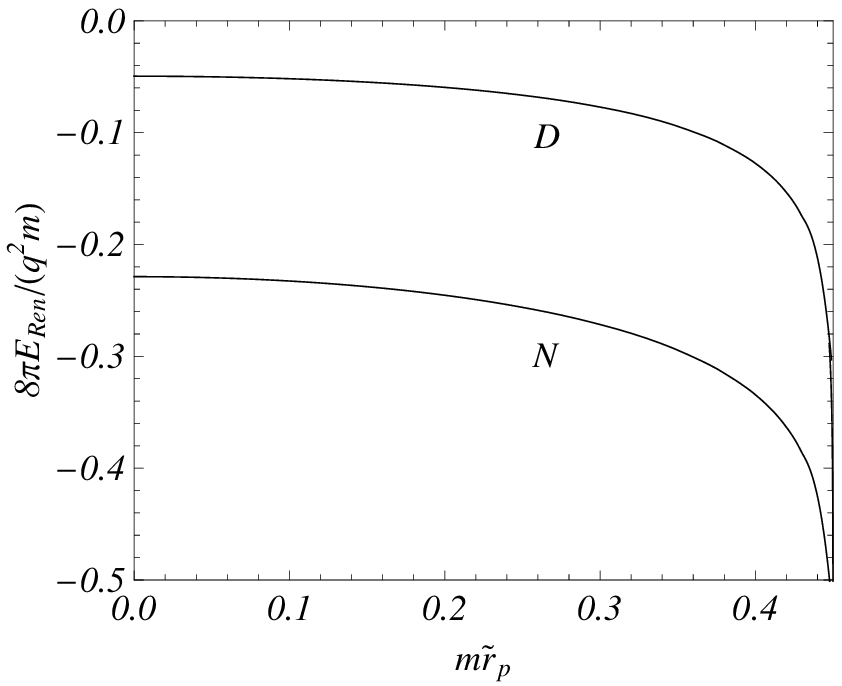, width=7.cm, height=6.5cm,angle=0} & \quad %
\epsfig{figure=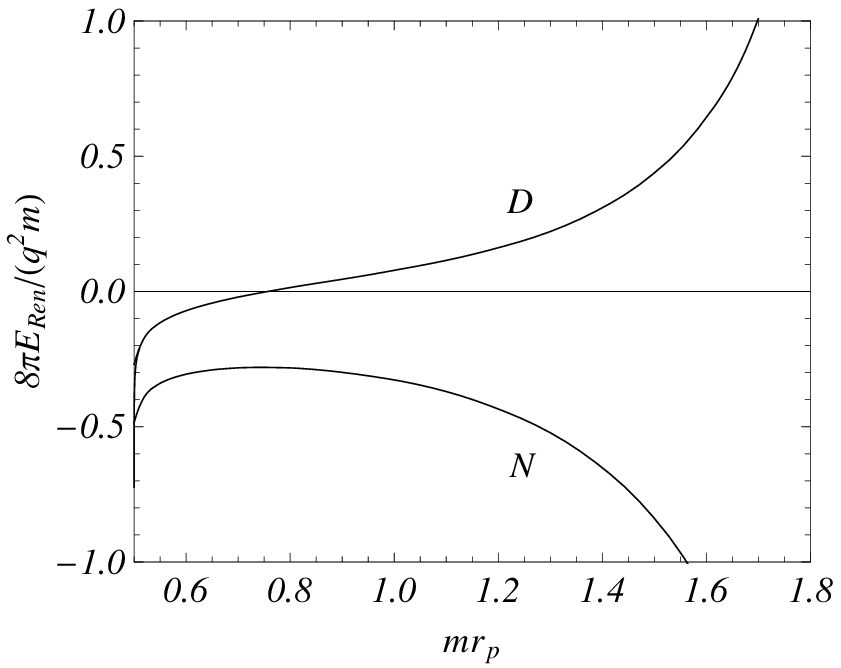, width=7.cm, height=6.5cm,angle=0}%
\end{tabular}%
\end{center}
\caption{The behavior of the self-energy in the flower-pot model as a
function of $m\tilde{r}_{p}$ and $mr_{p}$ considering the region inside the
core in the left panel and outside the core in the right panel. In both
cases we adopted $\protect\alpha =0.9$, $ma=1/2$, $mR=2$ and the graphs are
plotted for a minimally coupled field.}
\label{fig3}
\end{figure}

In figure \ref{fig3} we exhibit, for the flower-pot model, the behavior of
the renormalized self-energy for a charged test particle placed inside (left
plot) and outside (right plot) the monopole's core as a function of $m\tilde{%
r}_{p}$ and $mr_{p}$, respectively. In these analysis we consider minimal
coupling ($\xi =0$), $\alpha =0.9$, $ma=1/2$ and $mR=2$. These two plots
exhibit the logarithmic singular behavior for the self-energy near the
core's boundary; moreover, the left panel shows the influence of the
boundary condition obeyed by the field on the external spherical boundary.
\footnote{%
Another graphs not presented in this paper show that this influence
decreases for larger value of $mR$; specifically for $mR=4$ no essential
difference between the plots has been observed. This is an expected
behavior, being a consequence of the dominant contribution on the
self-energy due to the core's structure itself.}

\subsection{Ballpoint pen model}

\label{Ballpoint}

The second model that we shall consider is the ballpoint pen one. For this
model, in the region inside the core, we may use the line element below,
\begin{equation}
ds^{2}=-dt^{2}+v^{2}(r)dr^{2}+\alpha ^{2}r^{2}(d\theta ^{2}+\sin ^{2}\theta
d\varphi ^{2})\ ,  \label{gm2}
\end{equation}%
where the function $v(r)$ is given by the expression (\ref{BP}). The only
non-vanishing components of the Ricci tensor are:
\begin{equation}
R_{r}^{r}=R_{\theta }^{\theta }=R_{\varphi }^{\varphi }=2\frac{1-\alpha ^{2}%
}{\alpha ^{2}a^{2}}\ .
\end{equation}%
As to the Ricci scalar, it reads
\begin{equation}
\mathcal{R}=6\frac{1-\alpha ^{2}}{\alpha ^{2}a^{2}}\ .
\end{equation}

By using the coordinate system corresponding to \eqref{gm2}, the two
independent solutions for the homogeneous radial differential equation
associated with (\ref{gr}), compatible with the Wronskian normalization (\ref%
{Wronin}), are
\begin{equation}
R_{1l}(r)=\frac{1}{\sqrt{\alpha r}}P_{\chi }^{-l-1/2}(z(r))\ \ \mathrm{and}\
\ R_{2l}(r)=\frac{(-1)^{l}\pi }{2\sqrt{\alpha r}}P_{\chi }^{l+1/2}(z(r))\ ,
\label{RBP}
\end{equation}%
being $P_{\mu }^{\nu }(x)$ the associated Legendre function and
\begin{equation}
z(r)=\sqrt{1-(r/a)^{2}(1-\alpha ^{2})}\ \ ,\ \ \chi =-\frac{1}{2}+\sqrt{%
1-6\xi -\frac{(ma\alpha )^{2}}{1-\alpha ^{2}}}\ .
\end{equation}%
Having the radial functions, we are in position to obtain the self-energy
for the regions inside and outside the monopole core. Again, these
expressions depend on the coefficients $D_{l}^{(+)}(a)$ and $D_{l}^{(-)}(a)$.

Considering first the exterior region, the self-energy is given by %
\eqref{TE1}. For this case we have,
\begin{equation}
D_{l}^{(+)}(a)=-\frac{n_{l}(a)}{d_{l}(a)}\ ,  \label{DBP+-}
\end{equation}%
with
\begin{eqnarray}
n_{l}(a) &=&\sqrt{1/\alpha ^{2}-1}I_{\nu _{l}}(ma)P_{\chi }^{-l+1/2}(\alpha )
\notag \\
&&-\left[ (\nu _{l}+l+1/2)I_{\nu _{l}}(ma)+maI_{\nu _{l}+1}(ma)\right]
P_{\chi }^{-l-1/2}(\alpha ),  \label{nla}
\end{eqnarray}%
and
\begin{eqnarray}
d_{l}(a) &=&\sqrt{1/\alpha ^{2}-1}K_{\nu _{l}}(ma)P_{\chi }^{-l+1/2}(\alpha )
\notag \\
&&-\left[ (\nu _{l}+l+1/2)K_{\nu _{l}}(ma)-maK_{\nu _{l}+1}(ma)\right]
P_{\chi }^{-l-1/2}(\alpha ).  \label{dla}
\end{eqnarray}%
In deriving the expressions (\ref{nla}) and (\ref{dla}) we have used the
relation \cite{Grad}%
\begin{equation}
(1-x^{2})\frac{dP_{\chi }^{\mu }(x)}{dx}=-\sqrt{1-x^{2}}P_{\chi }^{\mu
+1}(x)-\mu xP_{\chi }^{\mu }(x).  \label{LegDer}
\end{equation}%
for the derivative of the associated Legendre function.

In \cite{Barbosa2} we have shown that, for this model, there is no divergent
contribution on the self-energy caused by the core-induced term for points
near the core; however, we may wonder about the influence of the core on the
boundary-induced term near the boundary. The relevant term to be analyzed
is,
\begin{equation}
E^{b}=\frac{q^{2}}{8\pi \alpha ^{2}r_{p}}\sum_{l=0}^{\infty }(2l+1)M_{\nu
_{l}}^{2}(mr_{p})\frac{\bar{K}_{\nu _{l}}(mR)}{\bar{M}_{\nu _{l}}(mR)}\ .
\end{equation}%
The procedure to investigate the behavior of this term is similar what we
have done in the last subsection. We have to analyze the term inside the
summation for large values of $l$. By using the uniform asymptotic
expansions for the modified Bessel functions, we obtain
\begin{equation}
D_{l}^{(+)}(a)\approx \frac{1}{\pi }\frac{R_{1l}(a)(2\nu _{l}-1)-2a{\mathcal{%
R}}_{l}^{(1)}(a)}{R_{1l}(a)(2\nu _{l}+1)+2a{\mathcal{R}}_{l}^{(1)}(a)}\left(
\frac{ema}{2\nu _{l}}\right) ^{2\nu _{l}}\ .
\end{equation}%
Considering the solutions \eqref{RBP}, the term inside the bracket, for
large values of $l$, reads,
\begin{equation}
\frac{R_{1l}(a)(2\nu _{l}-1)-2a{\mathcal{R}}_{l}^{(1)}(a)}{R_{1l}(a)(2\nu
_{l}+1)+2a{\mathcal{R}}_{l}^{(1)}(a)}\approx \frac{(1-\alpha )}{8l^{2}}%
[4\chi (1+\chi )-\alpha (1+\alpha )(1-8\xi )]\ .
\end{equation}%
So on basis of this result, near the boundary we may approximate $M_{\nu
_{l}}(z)\approx I_{\nu _{l}}(z)$, for $z=mr_{p}$ or $z=mR$. Consequently the
above expression can be written as,
\begin{equation}
E^{b}\approx \frac{q^{2}}{8\pi \alpha ^{2}r_{p}}\sum_{l=0}^{\infty
}(2l+1)I_{\nu _{l}}^{2}(mr_{p})\frac{\bar{K}_{\nu _{l}}(mR)}{\bar{I}_{\nu
_{l}}(mR)}\ ,
\end{equation}%
which coincides with the boundary-induced term given in \eqref{TE2} for the
point-like monopole.

The general formula for the self-energy in the region $a\leqslant
r_{p}\leqslant R$ is simplified for a massless scalar field. Taking the
limit $m\rightarrow 0$ in this formula, one finds%
\begin{equation}
E_{\mathrm{Ren}}=-\frac{q^{2}}{8\pi \alpha r_{p}}S_{(\alpha )}(0)+\frac{q^{2}%
}{8\pi r_{p}}\sum_{l=0}^{\infty }\frac{2l+1}{2\alpha ^{2}\nu _{l}}\left\{
e_{l}+\frac{\left[ (r_{p}/a)^{2\nu _{l}}-e_{l}\right] ^{2}}{F_{l}(R/a)^{2\nu
_{l}}-e_{l}}\right\} \left( \frac{a}{r_{p}}\right) ^{2\nu _{l}},
\label{Ebpm0}
\end{equation}%
where we use the notation%
\begin{equation}
e_{l}=\frac{\sqrt{1/\alpha ^{2}-1}P_{\chi _{0}}^{-l+1/2}(\alpha )-(l+1/2+\nu
_{l})P_{\chi _{0}}^{-l-1/2}(\alpha )}{\sqrt{1/\alpha ^{2}-1}P_{\chi
_{0}}^{-l+1/2}(\alpha )-(l+1/2-\nu _{l})P_{\chi _{0}}^{-l-1/2}(\alpha )},
\label{el}
\end{equation}%
with%
\begin{equation}
\chi _{0}=-\frac{1}{2}+\sqrt{1-6\xi }.  \label{xi0}
\end{equation}%
The core structure is encoded in the function $e_{l}$.

Let us now analyze the scalar self-energy for the region inside the core. We
need to substitute the functions (\ref{RBP}) into the formulas (\ref{G0in})
and (\ref{Galfain}) to construct the corresponding Green function, and in (%
\ref{G0renin}) to obtain the renormalized self-energy, Eq. (\ref{ERin}). So
we have:
\begin{equation}
G_{0}(\mathbf{x},\mathbf{x}^{\prime })=\frac{1}{8\alpha \sqrt{rr^{\prime }}}%
\sum_{l=0}^{\infty }(2l+1)(-1)^{l}P_{\chi }^{-l-1/2}(z_{<})P_{\chi
}^{l+1/2}(z_{>})P_{l}(\cos \gamma )\ ,  \label{G0inBP}
\end{equation}%
where
\begin{equation}
z_{\gtrless }=\sqrt{1-(r_{\gtrless }/a)^{2}(1-\alpha ^{2})}\ .  \label{X}
\end{equation}%
As to DeWitt-Schwinger expansion of the Green function we use (\ref{Had3D}).
Taking first the coincidence limit on the angular variables, the radial
geodesic distance between two close points inside the monopole can be
approximately given by
\begin{equation}
\sqrt{2\sigma }\approx \frac{\alpha |r^{\prime }-r|}{z(r)}\ .
\end{equation}%
So we get
\begin{equation}
G_{\mathrm{Sing}}(r^{\prime },r)=\frac{z(r)}{4\pi \alpha |r^{\prime }-r|}-%
\frac{m}{4\pi }\ .  \label{Gsing-BP}
\end{equation}

Developing some intermediate steps, we obtain the following expression for $%
G_{0,\mathrm{Ren}}(r^{\prime },r)$:
\begin{equation}
G_{0,\mathrm{Ren}}(r,r)=\frac{S_{\alpha }(r/a)}{4\pi \alpha r}+\frac{m}{4\pi
}\ ,  \label{Gr0}
\end{equation}%
where
\begin{equation}
S_{\alpha }(r/a)=\sum_{l=0}^{\infty }\left[ F\left( -\chi ,\ \chi +1;\
1/2-l;\ (1-z(r))/2\right) F\left( -\chi ,\ \chi +1;\ 3/2+l;\
(1-z(r))/2\right) -1\right]  \label{Gr11}
\end{equation}%
being $F(a,b;c;z)$ the hypergeometric function \cite{Grad}.\footnote{%
The explicit calculation of \eqref{Gr0} is given in \cite{Barbosa2}.}
Consequently the renormalized self-energy becomes,
\begin{equation}
E_{\mathrm{Ren}}=-\frac{q^{2}S_{\alpha }(r_{p}/a)}{8\pi \alpha r_{p}}-\frac{%
q^{2}m}{8\pi }-\frac{q^{2}}{16\alpha r_{p}}\sum_{l=0}^{\infty
}(2l+1)(-1)^{l}D_{l}^{(-)}(a)[P_{\chi }^{-l-1/2}(z(r_{p}))]^{2}\ .
\label{EintBP}
\end{equation}

Let us now analyze the influence of the boundary on the scalar self-energy
near the core's center. This can be done by calculating explicitly the
coefficient $D_{l}^{(-)}(a)$. By expressing the associated Legendre function
in terms of hypergeometric one by \cite{Grad},
\begin{equation}
P_{\chi }^{\mu }(x)=\frac{1}{\Gamma (1-\mu )}\left( \frac{1+x}{1-x}\right) ^{%
\frac{\mu }{2}}F\left( -\chi ,\ \chi +1;\ 1-\mu ;\frac{1-x}{2}\right) \ ,
\label{Legend}
\end{equation}%
we can see that taking $r\rightarrow 0$, we may approximate $z(r)\approx
1-(r^{2}/2a^{2})(1-\alpha ^{2})$, and by (\ref{Legend}) the leading term in
the associated Legendre function reads,
\begin{equation}
P_{\chi }^{-l-1/2}(z(r))\approx \frac{1}{\Gamma (l+3/2)}\left( \frac{r\sqrt{%
1-\alpha ^{2}}}{2a}\right) ^{l+1/2}\ .
\end{equation}%
So the dominant component in the core-induced term is for $l=0$. Considering
also the contribution due to the mass term in (\ref{EintBP}), we have%
\footnote{%
The contribution due to the background geometry goes to zero for $%
r_{p}\rightarrow 0$.}
\begin{equation}
E_{\mathrm{Ren}}\approx -\frac{q^{2}D_{0}^{(-)}(a)\sqrt{1-\alpha ^{2}}}{%
8\alpha \pi a}-\frac{q^{2}m}{8\pi }\ .
\end{equation}%
In this case we can also see that for $R\gg a$, the leading term in $%
D_{0}^{(-)}(a)$ does not depend on $R$.

For a massless field the general formula for the self-energy inside the core
is specified to%
\begin{eqnarray}
E_{\mathrm{Ren}} &=&-\frac{q^{2}S_{\alpha }^{(0)}(r_{p}/a)}{8\pi \alpha r_{p}%
}+\frac{q^{2}}{16\alpha r_{p}}\sum_{l=0}^{\infty }(-1)^{l}(2l+1)  \notag \\
&&\times \frac{F_{l}f_{l+1/2}^{(-)}(R/a)^{2\nu _{l}}-f_{l+1/2}^{(+)}}{%
F_{l}f_{-l-1/2}^{(-)}(R/a)^{2\nu _{l}}-f_{-l-1/2}^{(+)}}[P_{\chi
_{0}}^{-l-1/2}(z(r_{p}))]^{2},  \label{Ebpinm0}
\end{eqnarray}%
where the expression for $S_{\alpha }^{(0)}(r_{p}/a)$ is obtained from (\ref%
{Gr11}) by the replacement $\chi \rightarrow \chi _{0}$ and%
\begin{equation}
f_{\nu }^{(\pm )}=\sqrt{1/\alpha ^{2}-1}P_{\chi _{0}}^{\nu +1}(\alpha )-(\pm
\nu _{l}-\nu )P_{\chi _{0}}^{\nu }(\alpha ).  \label{fnu}
\end{equation}%
Before to finish this application we want to emphasize that, as in the case
of the region outside the core, there is no divergent result on the
self-energy given by core-induced part.

\begin{figure}[tbph]
\begin{center}
\begin{tabular}{cc}
\epsfig{figure=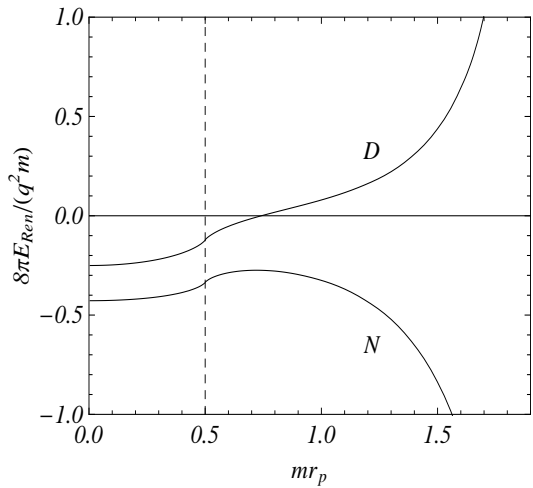, width=7.cm, height=6.5cm,angle=0} &
\end{tabular}%
\end{center}
\caption{The self-energy in the ballpoint pen model for a minmally coupled
scalar field as a function of $mr_{p}$, considering $\protect\xi =0$, $%
\protect\alpha =0.9$, $ma=1/2$ and $mR=2$.}
\label{fig4}
\end{figure}

In figure \ref{fig4}, for the ballpoint pen model we display the behavior of
the renormalized self-energy for a charged test particle placed inside and
outside the core as a function of $mr_{p}$. Also here we consider the
minimal coupling ($\xi =0$), $\alpha =0.9$, $ma=1/2$, and $mR=2$. For DBC
the corresponding self-force is repulsive with respect to the spherical
boundary. For NBC the force is repulsive with respect to the boundary for
points near the core's center and attractive for points near the boundary.

\section{Concluding remarks}

\label{Conc}

In this paper we have analyzed the renormalized self-energy and the
self-force associated with a static scalar charged particle placed in the
global monopole spacetime, considering a spherical boundary with radius $R$
larger than the core's radius, $a$. Moreover, on the scalar field associated
with the charge, we impose Dirichlet or Neumann boundary conditions on the
spherical boundary. The latter modify the self-energy introducing an extra
term, named $E^{b}$, when compared with the case in the absence of boundary.
The main part of this paper concerned on the investigation of the influence
of the boundary on this self-energy.

Considering first the particle outside the spherical boundary, we observed
that for points near it the self-energy is dominated by the boundary-induced
part, which presents a singular behavior given by \eqref{E-out}. The
corresponding self-force is repulsive/attractive with respect to the sphere
for DBC/NBC. At large distances from the sphere the self-force can be either
attractive or repulsive depending on the curvature coupling parameter. In
particular, for a minimally coupled scalar field the force is attractive for
both DBC and NBC. For the Dirichlet case, the self-energy has a minimum for
some intermediate value of the radial coordinate providing a stable
equilibrium point for a point particle.

The second part of the paper is devoted to the analysis of the self-energy
in the region inside the spherical boundary. Although at the beginning we
have developed a general formalism, only attributing specific models for the
region inside the monopole's core, some numerical and closed analysis can be
developed. Firstly we have considered a point-like global monopole. In this
model for points near the monopole the character of the force depends on the
curvature coupling parameter. In the special case of a minimally coupled
scalar field the self-force is attractive with respect to the monopole for
both DBC and NBC. For a monopole with non-trivial core structure, the force
depends on the specific model adopted. We have considered two specific
models, namely, the flower-pot and ballpoint pen models. For both these
cases we observed that the singular behavior of boundary-induced term in the
self-energy is not changed. This behavior is given by \eqref{ER+}. Besides,
in the opposite direction, we also proved that the presence of the boundary
does not affect the singular behavior of the core-induced contribution on
the self-energy for the flower-pot model. As to the self-energy near the
monopole's center, we could see that the presence of the boundary produces a
correction on the corresponding quantity when compared with the case of
absence of boundary; moreover, this correction becomes weaker for larger
values of the sphere radius. In the flower pot model, the force near the
core's boundary is attractive (repulsive) with respect to this boundary for $%
\xi <1/8$ ($\xi >1/8$) for both DBC and NBC. In the ballpoint pen model and
for a minimally coupled scalar field, the self-force is attractive with
respect to the core's center for the region inside the core and also for the
exterior points near the core's surface.

The complete analysis for the self-energies, in the regions inside and
outside the core, are exhibited numerically in the graphs \ref{fig3} and \ref%
{fig4} for the flower-pot model and ballpoint pen model, respectively. For
the flower-pot model, the graphs exhibit the singular behavior near the
core's boundary; as to the ballpoint pen one, we were able to provide the
behavior of the self-energy for both regions in a single plot.

\section*{Acknowledgment}

E.R.B.M. thanks Conselho Nacional de Desenvolvimento Cient\'{\i}fico e Tecnol%
\'{o}gico (CNPq) for partial financial support. A.A.S. was supported by the
Program PVE/CAPES.

\appendix

\section{Analysis of the self-energy for DBC and NBC}

\label{appA}

For a static field configuration, by using the field equation (\ref{EM}),
the $_{0}^{0}$-component of the energy-momentum tensor (\ref{Tmunu}) is
presented in the form
\begin{equation}
T_{0}^{0}=(\xi -1/4)\nabla ^{2}\phi ^{2}+\frac{1}{2}\rho \ \phi \ .
\label{Energy-density1}
\end{equation}%
Here we have used the fact that for the geometry under consideration $%
R_{0}^{0}=0$. The energy associated with the system under investigation is
given by substituting (\ref{Energy-density1}) into \eqref{Energy1}. It
reads,
\begin{equation}
E=-\int \ d^{3}x\ \sqrt{-g}\left[ (\xi -1/4)\nabla ^{2}\phi ^{2}+\frac{1}{2}%
\rho \ \phi \right] \ ,  \label{Energy2}
\end{equation}%
where the integral is evaluated for the region inside the boundary, $%
r\leqslant R$, or outside the boundary, $r\geqslant R$, for the cases of the
particle inside or outside the boundary, respectively.

Let us analyze the contribution of the integral below, present in %
\eqref{Energy2}:
\begin{equation}
\int \ d^{3}x\ \sqrt{-g}\ \nabla ^{2}\phi ^{2}.  \label{I1}
\end{equation}%
By applying the Gauss theorem, we can write this integral as,
\begin{equation}
\int \ d^{3}x\ \sqrt{-g}\ \nabla ^{2}\phi ^{2}=\int \ d^{3}x\ \sqrt{-g}\
\nabla _{i}V^{i}=\int \ d^{3}x\ \partial _{i}(\sqrt{-g}\ V^{i})=\oint \
dS_{i}\ \sqrt{-g}\ V^{i}\ ,  \label{I1b}
\end{equation}%
where we have introduced the three-vector $V^{i}=\nabla ^{i}\phi ^{2}$ and
used the fact that for static field $\nabla _{\mu }V^{\mu }=\nabla _{i}V^{i}$%
. Moreover, we have admitted that the field goes to zero at infinity. So,
the above integral only depends on the the field or its derivative
perpendicular to the boundary on the boundary itself. For the spherical
boundary we may write $dS_{i}=dSn_{i}$, where $n_{i}=(1,0,0)$ or $%
n_{i}=(-1,0,0)$ for the inner or outer regions, respectively. For DBC and
NBC, Eqs. (\ref{Diri1}) and (\ref{Neu1}), the surface integral vanishes and,
hence, the term with the integral (\ref{I1}) does not contribute to the
energy. Finally, the energy is presented in the form
\begin{equation}
E=-\frac{1}{2}\int \ d^{3}x\ \sqrt{-g}\ \rho \phi \ .  \label{Etot}
\end{equation}

Note that in the case of Robin boundary condition on the scalar
field the surface term (\ref{I1b}) does not vanish. However, it
should be taken into account that the energy-momentum tensor for a
scalar field on manifolds with boundaries in addition to the bulk
part contains a contribution located on the boundary. For an
arbitrary smooth boundary the expression for the
surface energy-momentum tensor is given in \cite{Saha04b} (see also \cite%
{Rome02} for the case of flat boundaries). By using this expression, it can
be seen that the part of the field energy coming from the surface
energy-momentum tensor exactly cancels the surface term in (\ref{I1b}) and
the energy is given by expression (\ref{Etot}) for Robin boundary condition
as well.

\section{Procedure adopted in the numerical evaluation}

\label{appB}

In this appendix we shall provide some technical details related with the
numerical analysis approach adopted by us. The most relevant numerical
analysis is related with the self-energy associated with the monopole
considered as a point-like defect. Because the numerical convergence of this
part is very delicate, it is convenient to adopt some trick, as it is
exhibited below. We consider the series given by (\ref{S}), where $\nu _{l}$
is given by \eqref{ind}. In order to improve the convergence, in the summand
we subtract and add the leading term in the asymptotic expansion for large
values of $l$. The latter is easily found by using the uniform asymptotic
expansions (\ref{UAE}):
\begin{equation}
I_{\nu }(z)K_{\nu }(z)\approx \frac{1}{2\nu }\left( 1-\frac{z^{2}}{2\nu ^{2}}%
\right) +O(\nu ^{-4})\ .
\end{equation}%
So, for the leading term we may write,
\begin{equation}
\frac{2l+1}{\alpha }I_{\nu _{l}}(mr)K_{\nu _{l}}(mr)-1\approx -\frac{%
(1-\alpha ^{2})(8\xi -1)+4\alpha ^{2}(mr)^{2}}{2(2l+1)^{2}}\ .
\end{equation}%
Finally, we may use in our numerical analysis the expression below:
\begin{eqnarray}
&&\sum_{l=0}^{\infty }\left[ \frac{2l+1}{\alpha }I_{\nu _{l}}(mr)K_{\nu
_{l}}(mr)-1\right] =\sum_{l=0}^{\infty }\left[ \frac{2l+1}{\alpha }I_{\nu
_{l}}(mr)K_{\nu _{l}}(mr)-1\right.  \notag \\
&&+\left. \frac{(1-\alpha ^{2})(8\xi -1)+4\alpha ^{2}(mr)^{2}}{2(2l+1)^{2}}%
\right] -\frac{\pi ^{2}}{16}[(1-\alpha ^{2})(8\xi -1)+4\alpha
^{2}(mr)^{2}]\ ,
\end{eqnarray}%
with improved convergence of the series.

\end{document}